\documentclass[acmsmall]{acmart}

\makeatletter                   
\def\mdseries@tt{m}             
\makeatother                    

\renewcommand\footnotetextcopyrightpermission[1]{} 
\usepackage{makecell}
\usepackage[plain]{fancyref}
\usepackage[draft=true]{minted}     

\newcolumntype{L}[1]{>{\raggedright\let\newline\\\arraybackslash\hspace{0pt}}m{#1}}
\newcolumntype{C}[1]{>{\centering\let\newline\\\arraybackslash\hspace{0pt}}m{#1}}
\newcolumntype{R}[1]{>{\raggedleft\let\newline\\\arraybackslash\hspace{0pt}}m{#1}}
\DeclareMathOperator*{\argmax}{arg\,max}

\setcopyright{none}
\acmDOI{10.3389/fphy.2018.00027}
\acmYear{2018}
\copyrightyear{2018}
\acmArticle{27}

\begin{document}
\sloppy   
\title{The New Urban Success: How Culture Pays}

\author{Desislava Hristova}
\affiliation{%
  \institution{Cambridge University}
  \city{Cambridge}
  \country{Cambridge, UK}
}
\email{desislava.hristova@cl.cam.ac.uk}

\author{Luca Maria Aiello}
\affiliation{%
  \institution{Nokia Bell Labs}
  \city{Cambridge}
  \country{Cambridge, UK}
	}
\email{luca.aiello@nokia.com}

\author{Daniele Quercia}
\affiliation{%
  \institution{Nokia Bell Labs}
  \city{Cambridge}
  \country{Cambridge, UK}
	}
\email{quercia@cantab.net}

\renewcommand{\shortauthors}{D. Hristova et al.}

\begin{abstract}
Urban economists have put forward the idea that cities that are culturally interesting tend to attract ``the creative class'' and, as a result, end up being economically successful. Yet it is still unclear how economic and cultural dynamics mutually influence each other. By contrast, that has been extensively studied in the case of individuals. Over decades, the French sociologist Pierre Bourdieu showed that people's success and their positions in society mainly depend on how much they can spend (their economic capital) and what their interests are (their cultural capital). For the first time, we adapt Bourdieu's framework to the city context. We operationalize a neighborhood's cultural capital in terms of the cultural interests that pictures geo-referenced in the neighborhood tend to express. This is made possible by the mining of what users of the photo-sharing site of Flickr have posted in the cities of London and New York over 5 years. In so doing, we are able to show that economic capital alone does not explain urban development. The combination of cultural capital and economic capital, instead, is more indicative of neighborhood growth in terms of house prices and improvements of socio-economic conditions. Culture pays, but only up to a point as it comes with one of the most vexing urban challenges: that of gentrification.
\end{abstract}

\keywords{culture, cultural capital, Pierre Bourdieu, hysteresis effect, Flickr}

\maketitle

Original paper published on Frontiers: \url{https://doi.org/10.3389/fphy.2018.00027}

\section{Introduction}

The French sociologist Pierre Bourdieu argued that we all possess certain forms of social capital. A person has, for example, symbolic capital (markers of prestige) and cultural capital (knowledge and cultural interests). These are forms of wealth that individuals bring to the ``social marketplace''. His work ultimately had the goal of testing what he called  \emph{`the differential association' hypothesis}~\cite{bourdieu1984distinction,grenfell2014pierre}. This states that  individuals with similar composition of capital are more likely to meet, interact, form relationships, have similar lifestyles and, as a result, be of the same social class. In his surveys of French taste, Bourdieu proved this to be the case. In so doing, he also found what he called the \emph{hysteresis effect}, which refers to any societal change that provides opportunities for the already successful to succeed further.  During times of change, individuals with more economic and cultural capital are the first to head to new (more advantageous) positions. A similar argument could apply to cities as well: a city constantly changes, and neighborhoods with more economic and cultural capital will be the first to head to new positions, contributing to the city's economic success. 

Such an argument has not been widely studied in the city context, yet it is behind most modern urban renewal initiatives inspired by  the `creative class' theory. This theory holds that cities with high concentrations of the creative class (e.g., technology workers, artists, musicians) show higher levels of economic development~\cite{florida2014rise}. The creative city as a planning paradigm supports creativity and culture by design, providing a direct link between cultural amenities, the quality of life, and economic development~\cite{glaeser2011triumph,jacobs1969economy, yencken1988creative}. However, cities and neighborhoods which are considered exemplars of creativity today are yet ridden with social and economic inequality~\cite{evans2009creative}. Cities such as San Francisco, New York, and London display a glaring gap between high- and low-income residents~\cite{florida2003cities}.  It is therefore interesting to explore the complex interplay between economic success and cultural creativity. The challenge is that it is hard to capture culture---all the more so at the scale of entire cities. We partly tackle that challenge by making two main contributions:

\begin{itemize}
\item We quantify neighborhood cultural capital from the pictures taken in both London and New York City over the course of five years.  To this end, we build the first `urban culture' taxonomy which contains words related to cultural activities and groups these words into nine categories.  We create this taxonomy by  proposing a semi-automated 5-step approach that uses both a top-down classification of the creative industries and a bottom-up crowd-sourced knowledge discovery from both  Wikipedia and Flickr.  We then select  picture tags that match these words. These tags come from approximately 10M geo-referenced  pictures in  London and New York which were posted on Flickr from 2007 to 2015. As a result, each neighborhood in the two cities is characterized by the fraction of picture tags that belong to each of the nine cultural categories. 

\item For the first time, we test  Bourdieu's \emph{hysteresis effect} in the city context. We find that urban development is well explained by a combination of cultural and economic capital. This combination allow us to successfully predict property values in New York and London neighborhoods (with $R^2=0.56$ and $0.81$, respectively).

\end{itemize}

\section{Methods}

\subsection{Taxonomy of Urban Culture}

\begin{figure}[t!]
    \centering
    \includegraphics[width=0.6\textwidth]{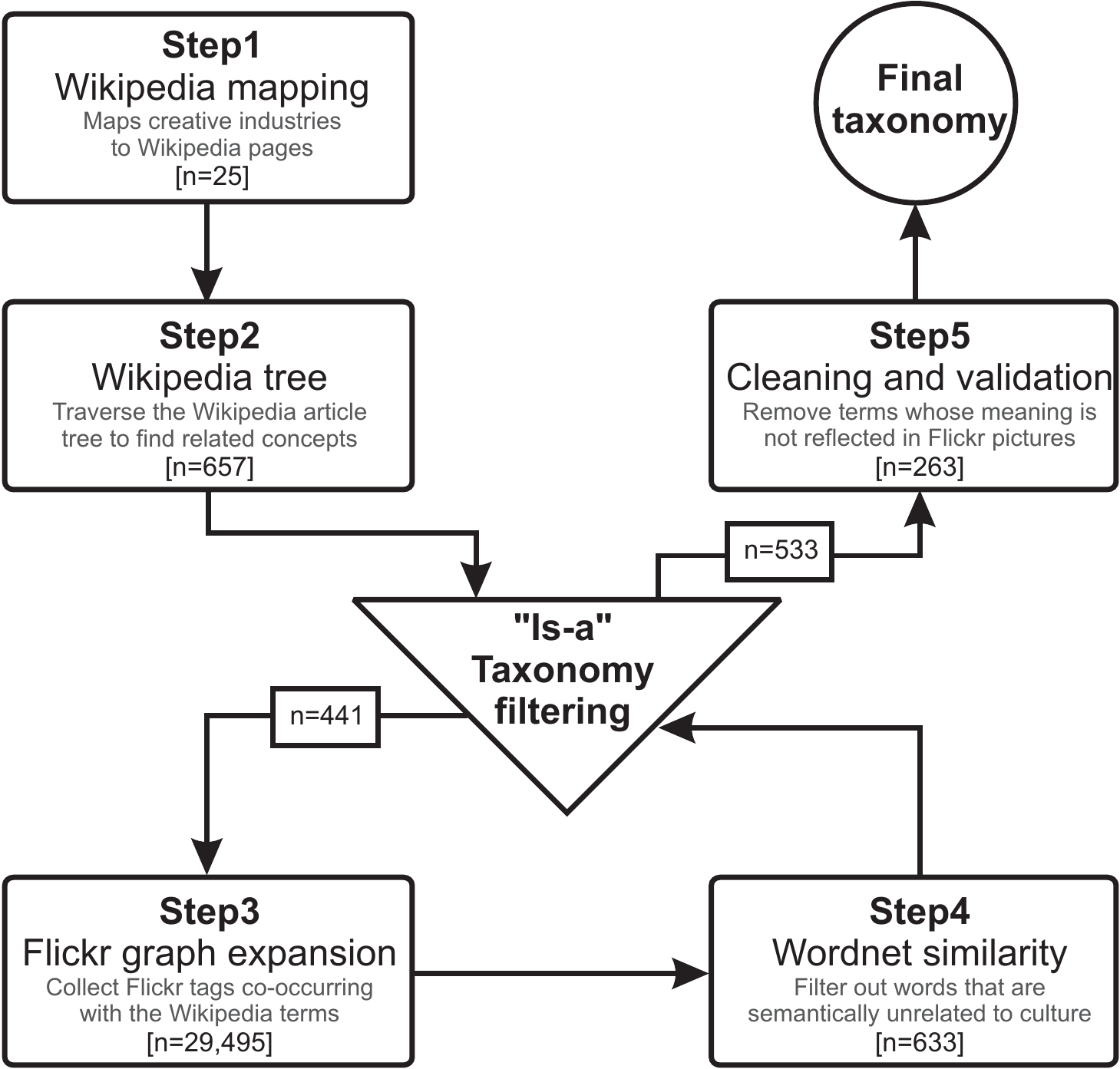}
    \caption{The five steps performed to obtain our  taxonomy of cultural terms}
    \label{fig:dia}
\end{figure}

In the urban setting, culture is mainly produced through the cultural services and artifacts of the creative industries. In 2015, the UK Department of Culture, Media and Sports adopted one of the most robust definitions of creative industries~\cite{bakhshi2012dynamic}. This  includes nine macro-industries:
\begin{itemize}
\item (100) Advertising and marketing 
\item (200) Architecture 
\item (300) Crafts 
\item (400) Design: product, graphic and fashion 
\item (500) Film, TV, video, radio and photography 
\item (600) IT software and computer services 
\item (700) Publishing
\item (800) Museums, galleries and libraries 
\item (900) Music, performing and visual arts
\end{itemize}
We then needed to expand this coarse-grained categorization into a multi-level taxonomy whose top nodes were these nine categories, intermediate nodes were subcategories, and leaf nodes were terms related to culture. Both subcategories and terms were to be determined, and we did so in five steps, which are described next (Figure~\ref{fig:dia}). 

\paragraph*{Step 1: Wikipedia Mapping}
A way to iteratively expand the initial nine categories is to connect them to an existing knowledge base of linked concepts. Because of its well-structured and hierarchically organized content, Wikipedia was fit for purpose, so much so that it had often been used to build semantically related large-scale taxonomies~\cite{ponzetto2007deriving}. However, each of the nine categories was hard to map to a single Wikipedia page. To ease the mapping, we disaggregated the nine categories based on their Wikipedia definitions. For example, we split `(400) Design' into each of the elements defined in its description: `Product Design', `Graphic Design' and `Fashion Design'.  After doing so for all the nine categories, we obtained twenty five  Wikipedia (Article) categories: (101) Advertising; (102) Marketing ; (200) Architecture; (300) Crafts; (400) Design; (401) Product design; (402) Graphic design; (403) Fashion; (501) Film; (502) Television; (503) Video; (504) Radio; (505) Photography; (601) Technology; (602) Gaming; (603) Software; (700) Publishing; (801) Arts; (802) Culture; (803) Museums; (804) Libraries; (901) Music; (902) Performance art; (903) Theatre; (904) Visual arts. We call these the \textit{top-level} categories.

\paragraph*{Step 2: Wikipedia Article Tree}
To expand these top-level categories, we use the Wikipedia graph. In general, Wikipedia category structure is essentially a graph of pages that can be navigated to find concepts that are related to each other. Starting from the twenty five top-level categories, we collected all the pages that \emph{directly} link to them (that is, those that are 1-hop distance apart in the graph\footnote{We do not navigate the graph at further hops because the number of connected pages grows exponentially at each hop, quickly including concepts that are highly unrelated.}). After automatically removing  community pages (which are not actual articles\footnote{Community pages are not Wikipedia articles. Instead, they belong to the following Wikipedia categories: \textsc{wikipedia, wikiprojects, lists, mediawiki, template, user, portal, categories, articles, images}}) and manually removing pages corresponding to highly ambiguous terms such as `color', we were left with 657subcategories (connected to the initial 25 top-level categories). 
  
\paragraph*{First `is-a' filtering}
Not all the 657 subcategories are relevant. Our goal was to build a taxonomy. By definition, a taxonomy connects categories and subcategories that are related with `is-a' relationships (e.g., the subcategory`film' is-a `product'). The `is-a' relationships relevant to culture had been identified by Gunnar Tornqvist in his book ``The Geography of Creativity''~\cite{tornqvist2011geography} and were represented by what he called the 4-$P$s: $\{process,place,person,product\}$. Therefore, out of the 657 subcategories, we filtered away those that were not $\{process,place,person,product\}$ and kept  the remaining 441 \textit{subcategories} (e.g., \textsc{Architect} and \textsc{Buildings} are subcategories of \textsc{Architecture}). These subcategories form the second level of our taxonomy.

\begin{figure}[t!]
   \centering
	 \includegraphics[width=0.32\textwidth]{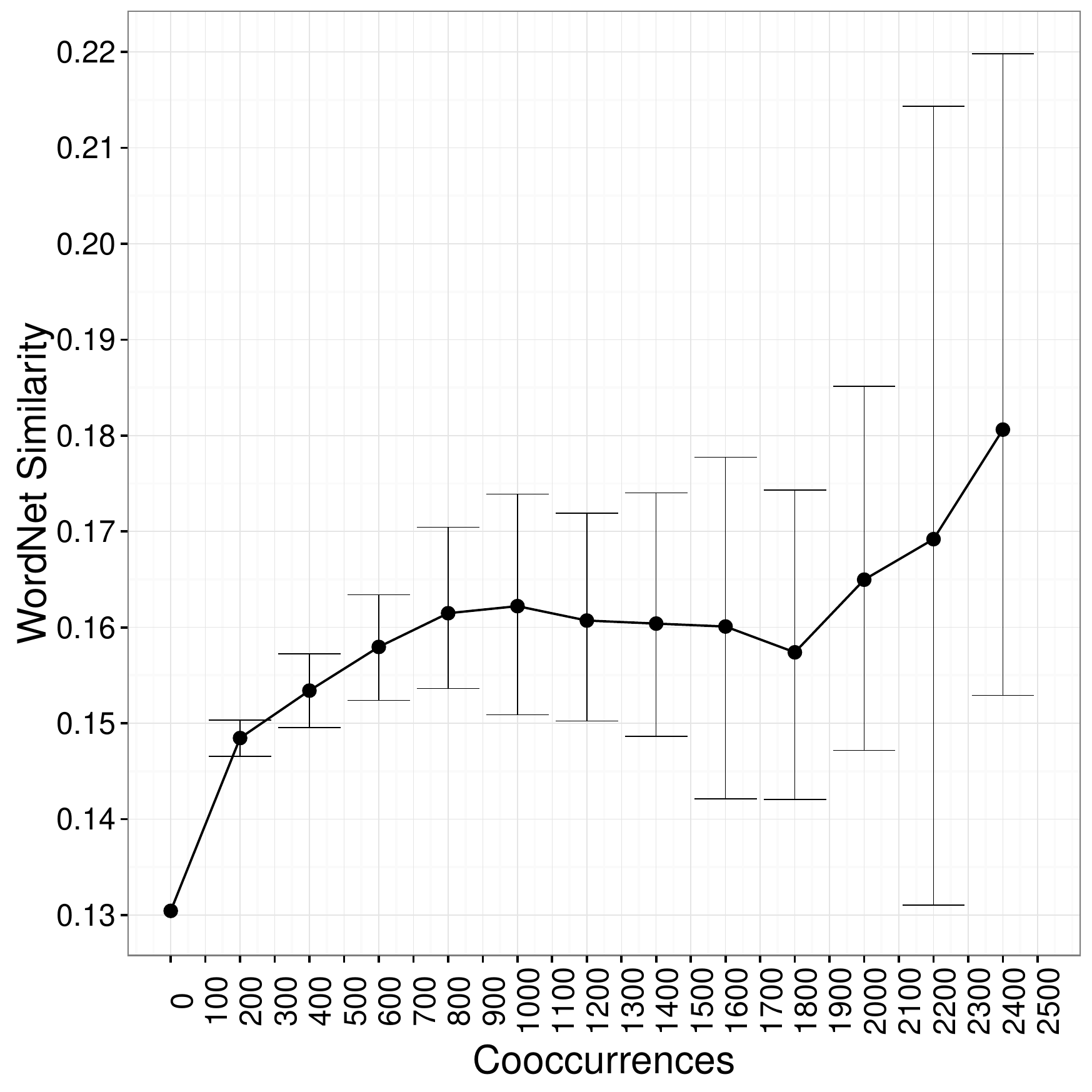} 
	 \includegraphics[width=0.32\textwidth]{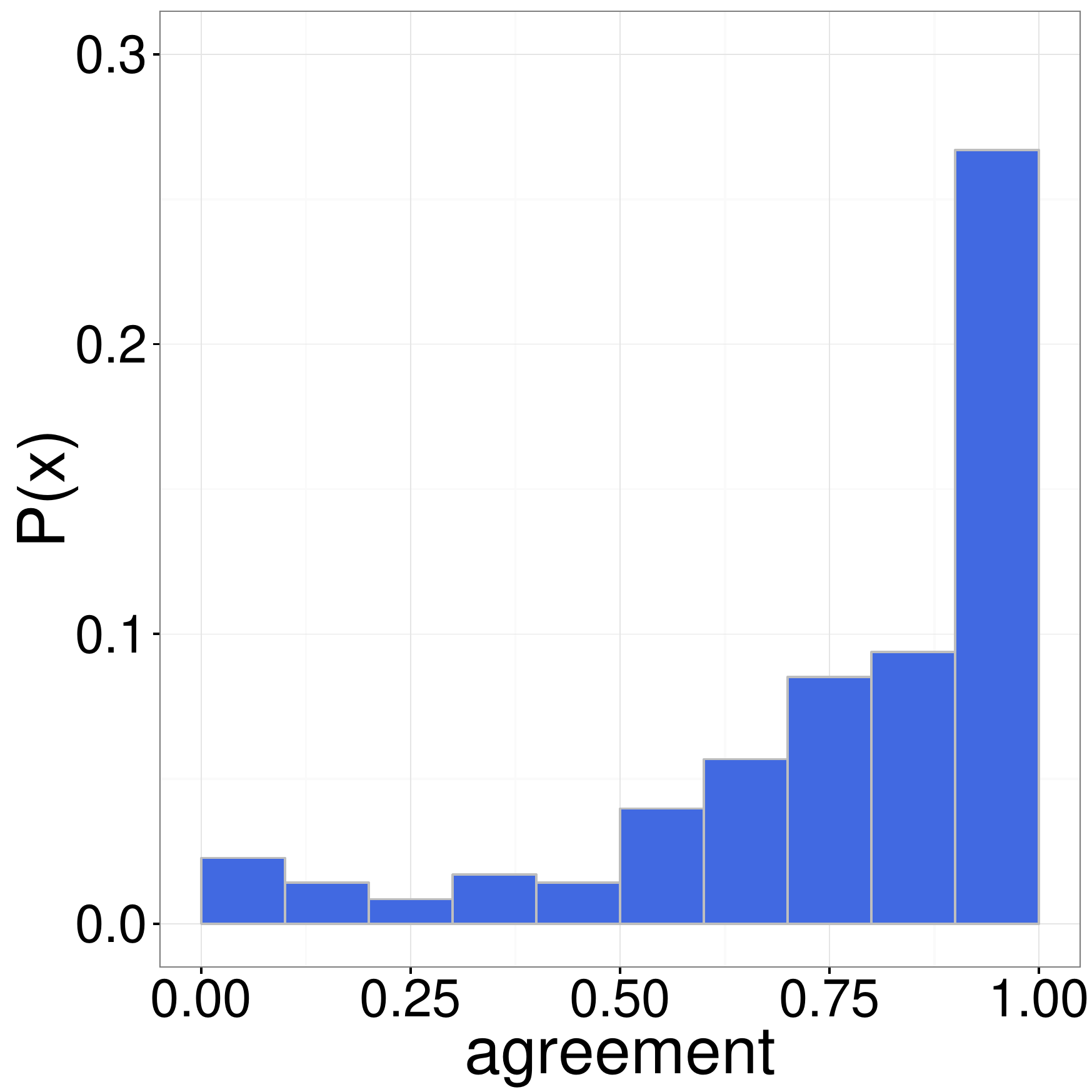} 
	\includegraphics[width=0.32\textwidth]{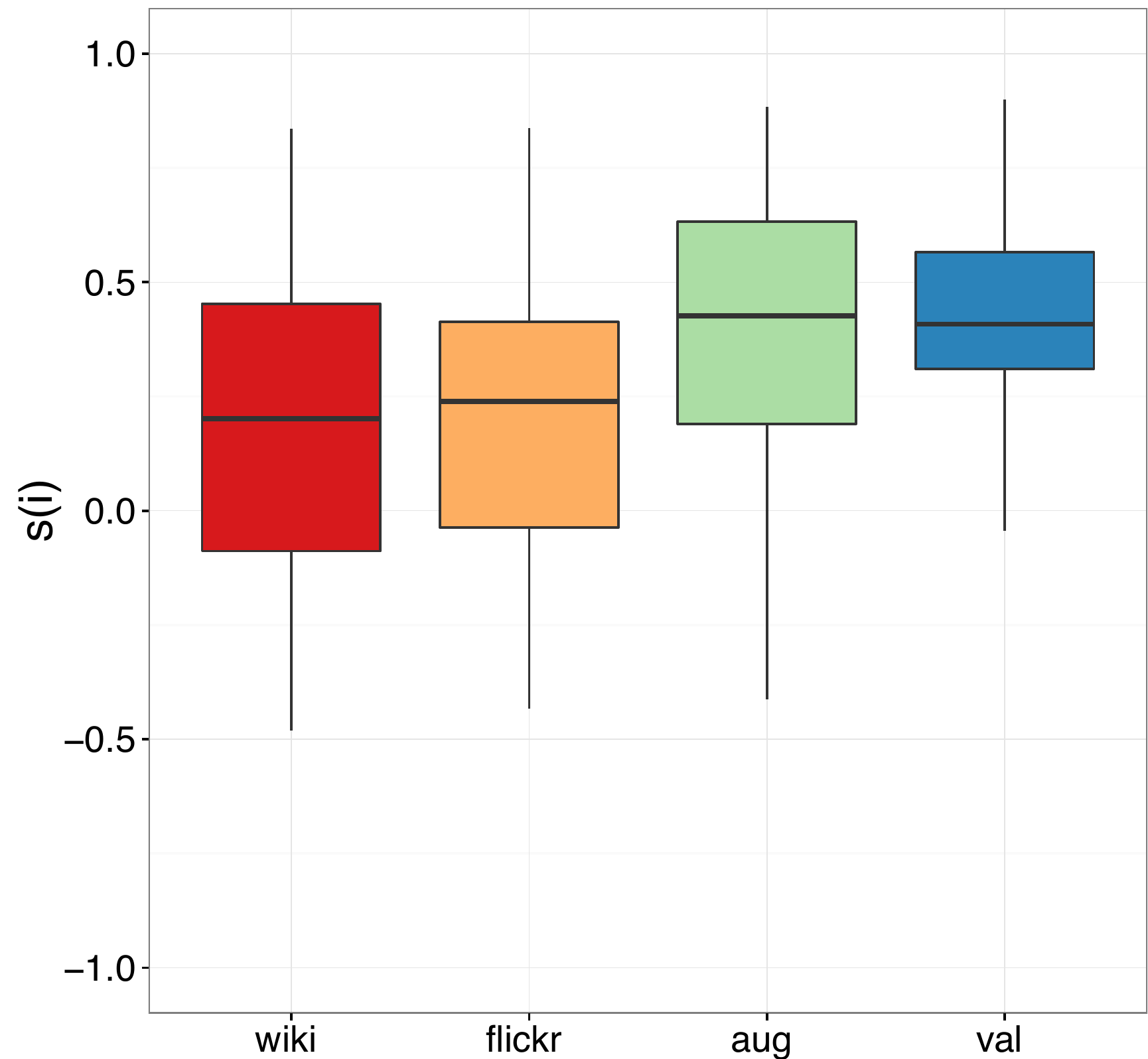} 
   \caption{\small (a) The average WordNet similarity for word pairs ($y$-axis) as the number of pair co-occurrences ($x$-axis) increases; (b) Agreement scores between pictures and cultural terms; (c) The silhouette value (the ``goodness'' of our taxonomy) at each creation step:  from the second step with the Wikipedia taxonomy only (\emph{wiki}), to the third with the Flickr graph expansion (\emph{flickr}), to the fourth which merged Flickr and Wikipedia (\emph{aug}), to the fifth which produced the validated and final taxonomy (\emph{val}). Each box shows the four quartiles of the distributions: the vertical lines indicate the top and bottom quartiles of values; the boxes are the mid-upper and lower quartiles, while the horizontal line in the middle shows the median value for the distribution. Outliers are shown as points on the graph.}
   \label{fig:pipeline_analysis}
\end{figure}

\paragraph*{Step 3: Flickr Graph Expansion}
To expand the taxonomy coverage as much as possible, we extended it with a third level containing specific terms related to the 441 subcategories. To do it with a data source complementary to Wikipedia, we relied on exploiting the structure behind tag co-occurrences on Flickr pictures. We did so because past studies had shown that tags that often co-occur in the same photo are semantically related to each other~\cite{quercia2015smelly}. We identified all the Flickr photos that contain at least one of the 441 terms, paired them with all the co-occurring Flickr tags, and characterized each pair with the corresponding number of  co-occurrences. By doing so, we found 373,849  co-occurrences: our  441 terms co-occurred with 29,495 new unique tags.

\paragraph*{Step 4: WordNet Similarity Filtering}
Of course, most of those co-occurrences were semantically irrelevant. We discarded the irrelevant ones by removing all pairs of terms that occurred a number of times less than a given threshold. To determine that threshold, we computed the \emph{similarity of term pairs} as a function of different co-occurrence thresholds: from a number of co-occurrences of 100 to one of 2500 (Figure~\ref{fig:pipeline_analysis}a). The similarity of a pair of terms  $t_1, t_2$ was computed using WordNet~\cite{fellbaum1998wordnet}:
\begin{equation}
sim_{path}(t_1, t_2) = 2*depth - len(t_1, t_2)
\label{eq:path}
\end{equation}
where $len(t_1, t_2)$ is the shortest path distance between $t_1$ and $t_2$ in WordNet, and $depth$ is the maximum distance between any two WordNet words.  The higher the value, the more similar the two terms. As done in previous work~\cite{meng2013review}, we computed the average path similarity ($\overline{sim}_{path}$) between all pairs of terms retained for each threshold value. In Figure~\ref{fig:pipeline_analysis}a, we see that the similarity considerably grows at first and then reaches a plateau at around a threshold value of 1800 co-occurrences. After it, the similarity still grows, but the corresponding standard deviations are too high. Therefore, conservatively, we kept all term pairs retained after applying a threshold of 2000 co-occurrences ($n=633$). 

\paragraph*{Second `is-a' filtering}
Not all the resulting 633 terms might have been related with `is-a' relationships, as the construction of a taxonomy would require. To double check that, we explored each of these 633 terms and manually filtered out those that were not linked with `is-a' relationships. This second `is-a' filtering resulted in 533 terms.

\begin{figure}[t!]
    \centering
    \includegraphics[width=0.60\textwidth]{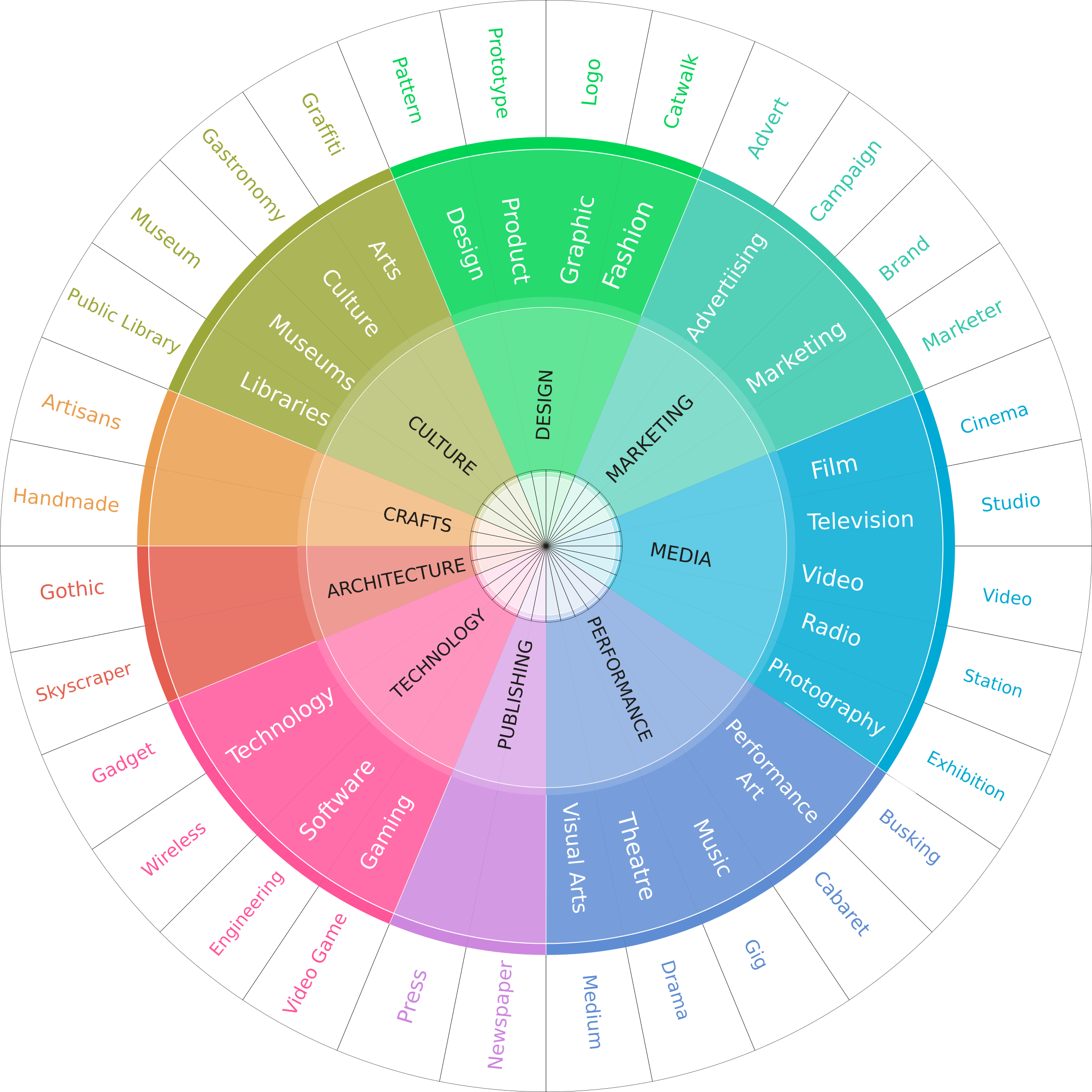}
    \caption{The wheel of our cultural taxonomy. The outer part shows examples of cultural terms (among the 263), the inner part shows the main 9 categories, and the middle part shows the 25 subcategories.}
    \label{fig:wheel}
\end{figure}

\paragraph*{Step 5: Cleaning}
To make sure that all the 533 terms were relevant, we performed a final cleaning step of potentially noisy terms. For each term, we drew a stratified random sample ($n=50$) of pictures marked with that term. We then labeled each image as either being related to the term or not. This made it possible to compute the average term's ``agreement'' with its corresponding 50 photos. We found that the majority of terms were in complete agreement with their photos and did reflect cultural assets (Figure~\ref{fig:pipeline_analysis}b). Conservatively, as a final step, we removed the terms that had an agreement lower than 0.75. This resulted in 263 terms, which are the leaf nodes of our final three-level taxonomy (Figure~\ref{fig:wheel}). 

\paragraph*{Validation}
The assumption behind the 5-step process was that each step resulted in a new set of terms that were better than the previous step's set. To ascertain whether that assumption was true, we measured whether the set of terms in the same top-level category (we have nine of such categories) was cohesive (the terms in the same category were all related to each other) and distinctive (the terms in different categories were orthogonal to each other). To that end, we measured the \textit{clustering silhouette}~\cite{rousseeuw1987silhouettes}. The silhouette $s$ of a term $i$ within cluster $C$ (its top-level category) determines how well $i$ lies within $C$:
\begin{equation}
s(i) = \frac{sim_{int}(i)-sim_{ext}(i)}{max\{sim_{int}(i),sim_{ext}(i)\}}
\label{eq:silhouette}
\end{equation}
where $sim_{int}(i)$ is the average path similarity (as per Formula~\ref{eq:path}) between term $i \in C$ and any other term $j$ that is in the \emph{same} cluster $C$; conversely, $sim_{ext}(i)$ is measured by first computing, for each cluster $C' \neq C$, the average similarity values between $i \in C$ and all the terms in $C'$ and then selecting the highest average similarity. The values of $s(i)$ range in [-1,1]: high values indicate cohesion, and low values indicate separation. 

We compared the distribution of silhouette values computed at each of the steps of  taxonomy creation (Figure~\ref{fig:pipeline_analysis}c): from step 2 (\emph{wiki}) to step 5 (\emph{val}). From the plot in Figure~\ref{fig:pipeline_analysis}c, we see that the median silhouette value indeed increases at each step: the median silhouette increases from 0.20 at step 2 (where only Wikipedia terms are considered) to 0.40 in the final step.

\begin{figure}[t!]
   \centering
   \includegraphics[width=0.30\textwidth]{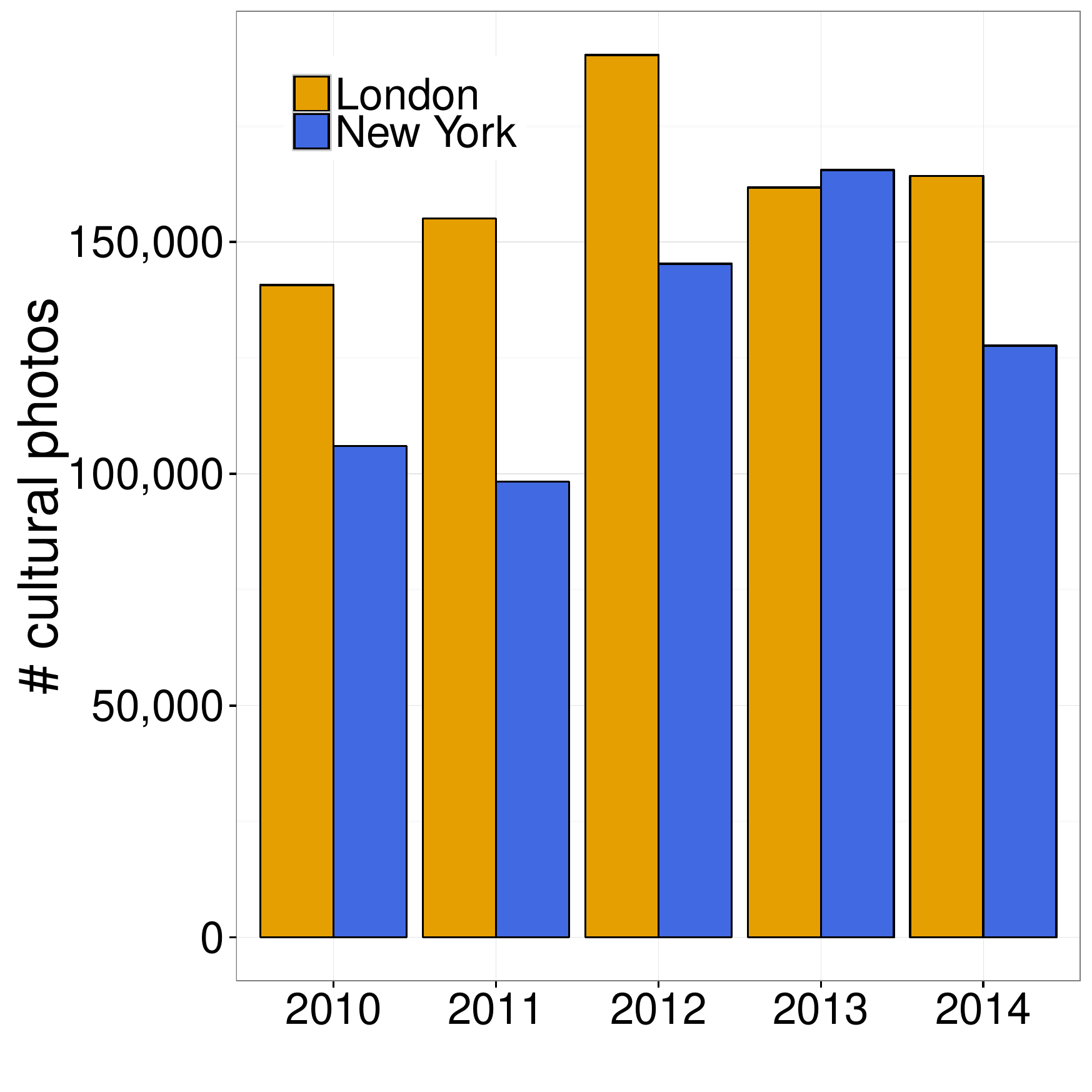} 
	 \includegraphics[width=0.30\textwidth]{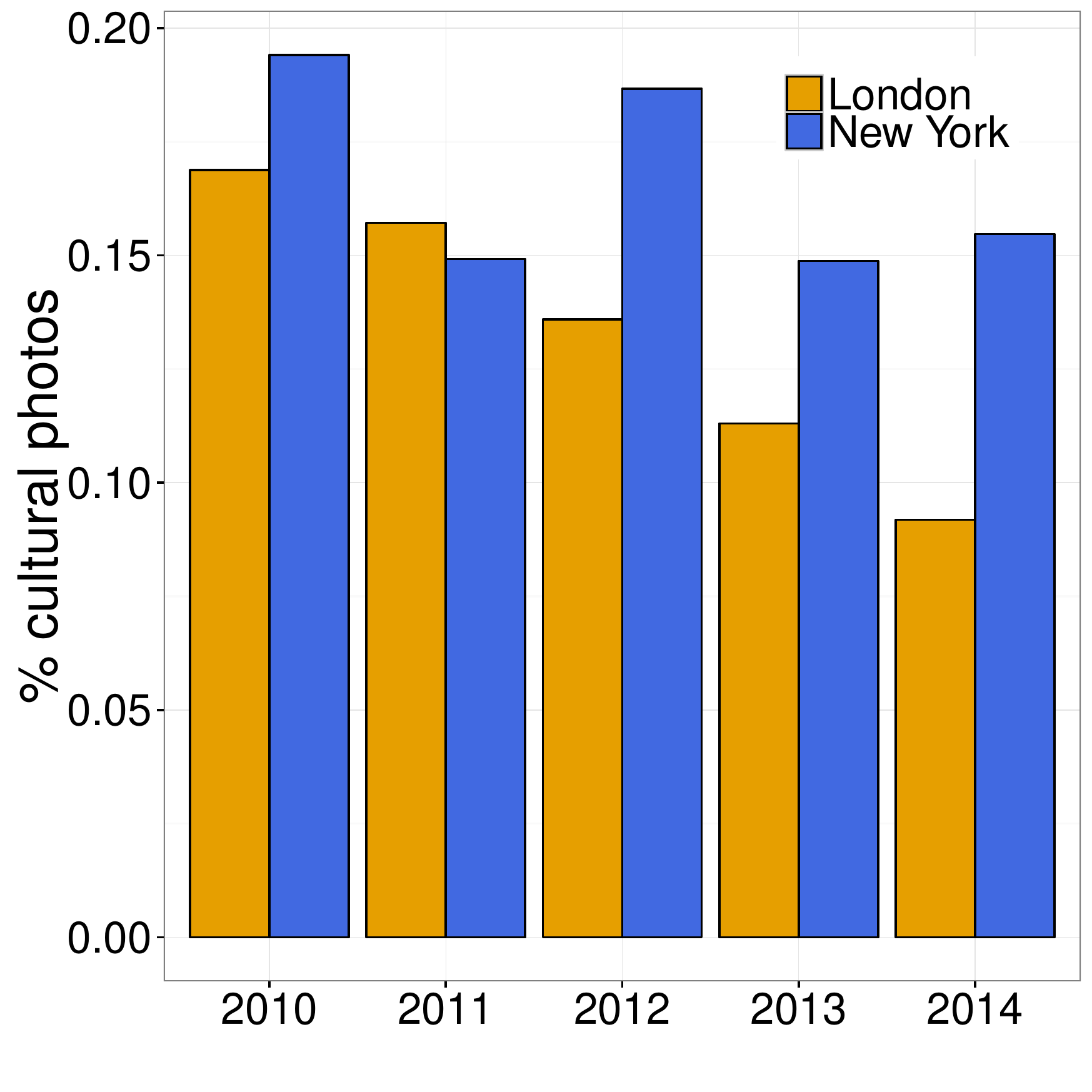} 
   \caption{Cultural content is consistently present over the five years under study: photos per annum (left) and fraction per annum (right).}
   \label{fig:dataset}
\end{figure}

\begin{figure}[t!]
    \centering
    \includegraphics[width=0.45\textwidth]{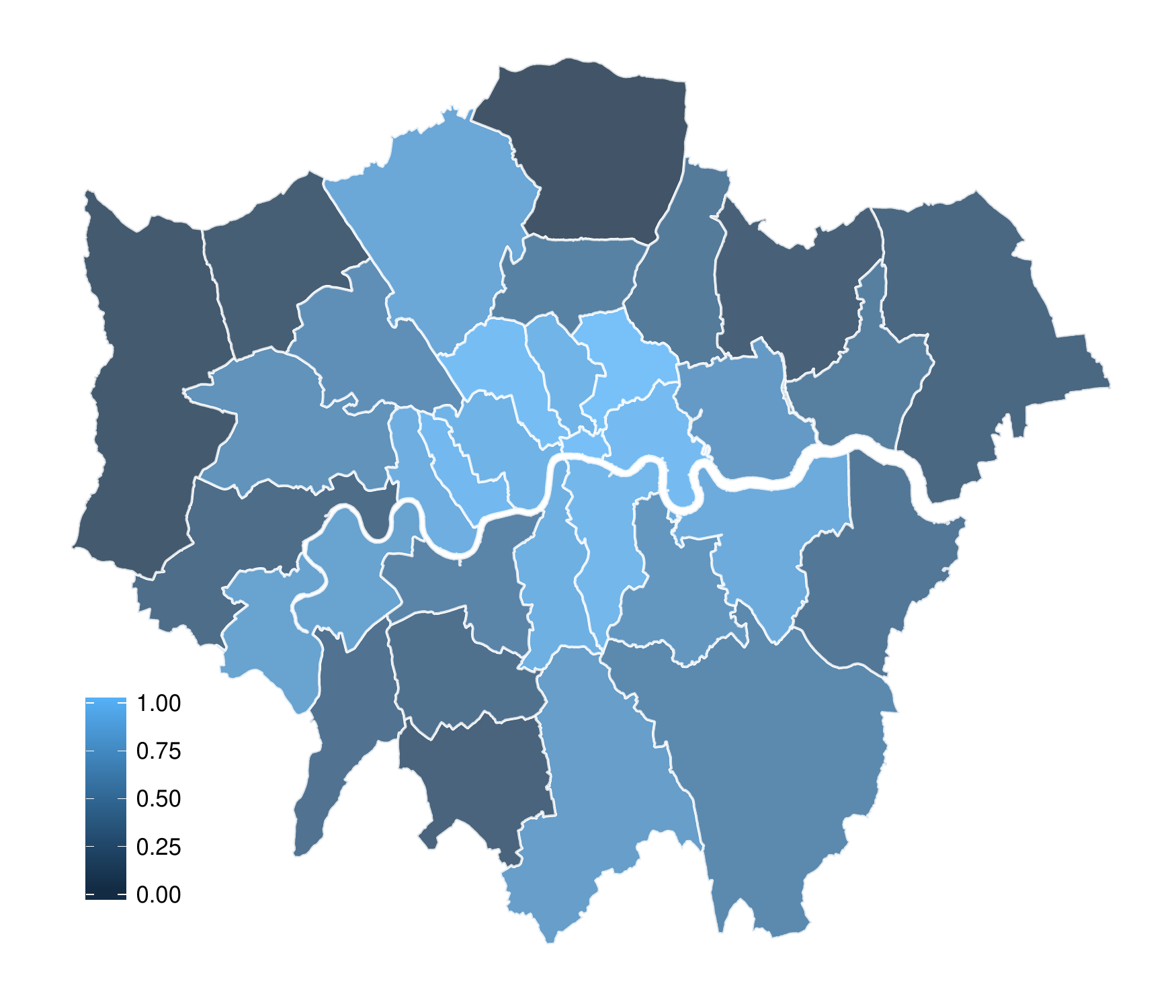}
		\includegraphics[width=0.45\textwidth]{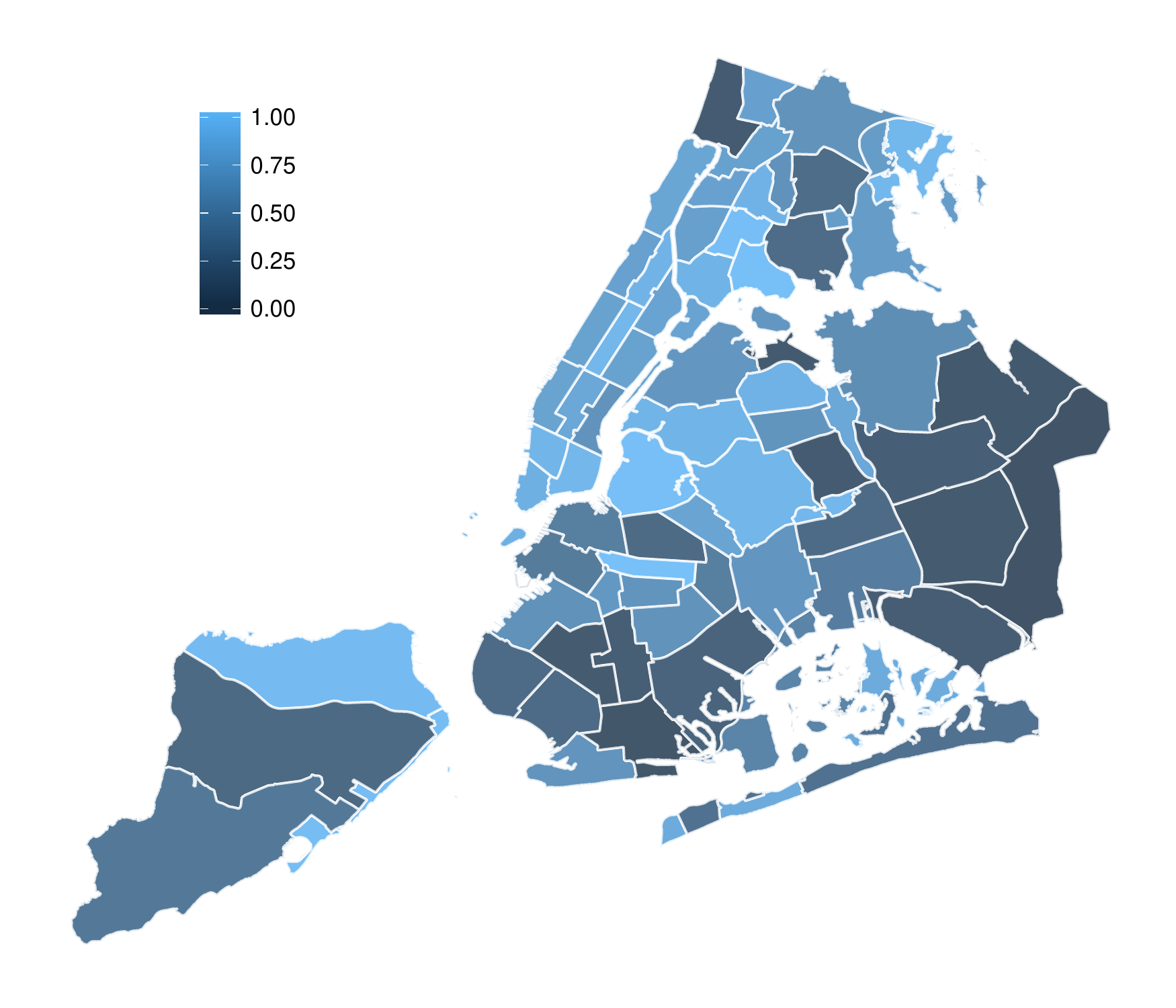}
    \caption{Cultural capital for neighborhoods in London (left) and New York (right). Neighborhoods are colored in terms of the amount of cultural capital they possess. The top 25\% of neighborhoods are depicted in light blue, while the bottom are the darkest.}
    \label{fig:cc}
\end{figure}

\subsection{Mapping cultural and economic capital}

Photos have been found to be good data sources for measuring people's perceptions of public places and for identifying distinctive features of the urban space (e.g., street art, temporary fairs) that are surveyed neither by the census nor by open mapping tools~\cite{aiello2016chatty, hristova2016measuring, manovich2009cultural, noulas2011exploiting, quercia2014shortest,quercia2015smelly}.  To trace cultural patterns in our user-generated  pictures, these pictures needed to be mapped onto geographical areas of interest. For London, we used its 33 boroughs; similarly, for New York, we used its 71 community districts (60 of which qualified for our analysis due to lack of data for the others). We assigned each picture to the corresponding census location $l$. To minimize the bias of our cultural profiling of cities towards amenities that are popular mostly among tourists, we filtered out non-locals by excluding any Flickr user who had been active in each of the two cities for less than 30 days, in a way similar to previous work~\cite{girardin2008digital}. We then retained only pictures marked with at least one tag matching one of the terms in our cultural taxonomy. This left us with $1.5M$ pictures. These pictures  covered  the period of five years with striking consistency  (Figure~\ref{fig:dataset}). They also  captured cultural vitality across neighborhoods. To see why, consider that, as opposed to New York, in London, the official number of cultural venues by borough is made publicly available~\footnote{Physical Asset Data. \url{https://www.gov.uk/government/statistical-data-sets/regional-and-local-insights-data}}. We correlated the number of our cultural-related pictures with the number of cultural venues  and found a Pearson correlation coefficient as high as 0.70 ($p < 0.01$).  

Following a methodology whose validity has been established in previous work~\cite{aiello2016chatty,quercia2015smelly}, to estimate the presence of cultural assets in each census location $l$, we computed the fraction of tags that match any of the words in our taxonomy:
\begin{equation}
f_{cult}(l) = \frac{\text{\# cultural tags @ } l}{\text{\# tags @ } l}.
\end{equation}
We then normalized those fractions using $z$-scores to obtain our estimate of \textit{cultural capital} for location $l$:
\begin{equation}
capital_{cult}(l) =  \frac{f_{cult}(l) - \mu(f_{cult})}{\sigma(f_{cult})},                                    
\end{equation}
where $\mu$ and $\sigma$ are, respectively, the mean and standard deviation of the $f_{cult}$ distribution over all locations. Values of $capital_{cult}$ are displayed on the map in Figure~\ref{fig:cc}. The values below zero indicate locations with fewer cultural activities than those in the average location, while values above zero indicate locations with greater cultural activities. Similarly, we computed an estimate of the \textit{economic capital} of a location as:
\begin{equation}
capital_{econ}(l) =  \frac{income(l) - \mu(income)}{\sigma(income)},                                    
\end{equation}
where $income(l)$ is the median income of resident taxpayers\footnote{Taxpayer income in London from the London Datastore:~\url{https://data.london.gov.uk/dataset/average-income-tax-payers-borough}; taxpayer income in New York from the American Community Survey:~\url{https://www.census.gov/programs-surveys/acs/data.html}}. The $z$-score represents the relatively high or low culture (or economic capital) that is characteristic of a location in units of standard deviations from the mean of the entire city. This allows us to draw an effective comparison not only between areas but also between the two forms of capital. To estimate the cultural capital under a specific top-level taxonomical category $c$, we computed the relative presence of tags of that top-level category in the location of interest, normalized across locations:
\begin{equation}
f_{cult}(l,c) = \frac{\text{\# cultural tags in category } c \text{ @ } l}{ \text{\# cultural tags @ } l};
\end{equation}
\begin{equation}
capital_{cult}(l,c) =  \frac{f_{cult}(l,c) - \mu(f_{cult}(c))}{\sigma(f_{cult}(c))}.                     
\end{equation}
To mark locations with their most distinctive type of cultural asset, we defined the \textit{cultural specialisation} of a location $l$ as the category $c$ with the highest capital:
\begin{equation}
special_{cult}(l) =  \argmax_{c}(capital_{cult}(l,c)).
\label{eq:specialisation}           
\end{equation}
To capture how diverse a location is in terms of the variety of dimensions expressed through cultural content, we computed its \textit{cultural diversity} as the Shannon entropy over the category-specific cultural capital values within that location:
\begin{equation}
diversity_{cult}(l) = H_{cult}(l) = - \sum_c f_{cult}(l,c) \times ln(p_{cult}(l,c)).
\label{eq:diversity}  
\end{equation}
Shannon Entropy does not take into account the finite size of the sample: low sample sizes (with respect to the number of bins) create biases towards higher entropy values. To overcome this problem we apply the Miller-Madow's correction to the entropy computation~\cite{paninski2003estimation}. High diversity values indicate locations with cultural interests that span across the nine top-level categories, while low diversity values indicate locations with cultural interests specialized in a specific top-level category.

\begin{figure*}[t!]
   \centering
   \includegraphics[width=0.23\textwidth]{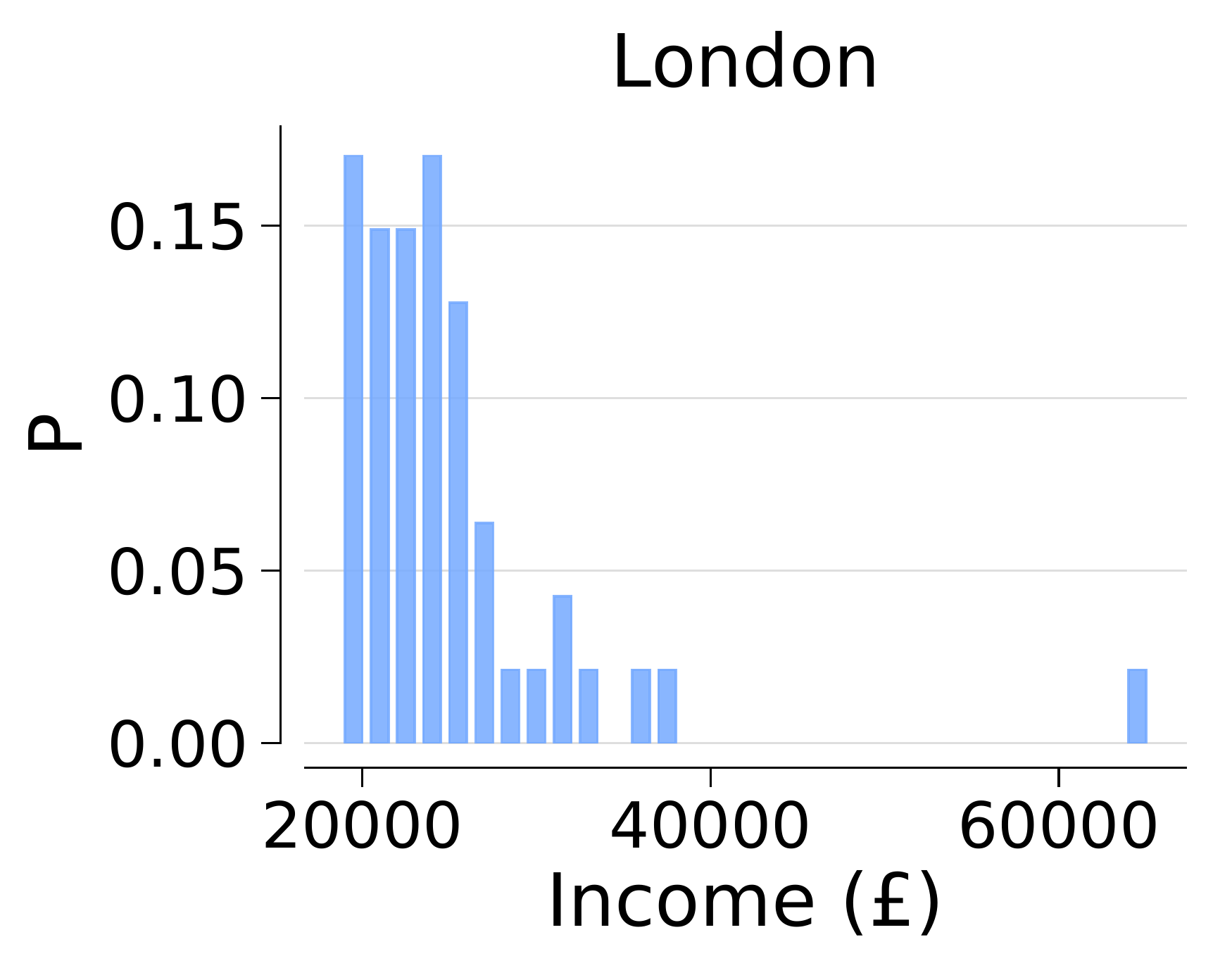} 
	 \includegraphics[width=0.23\textwidth]{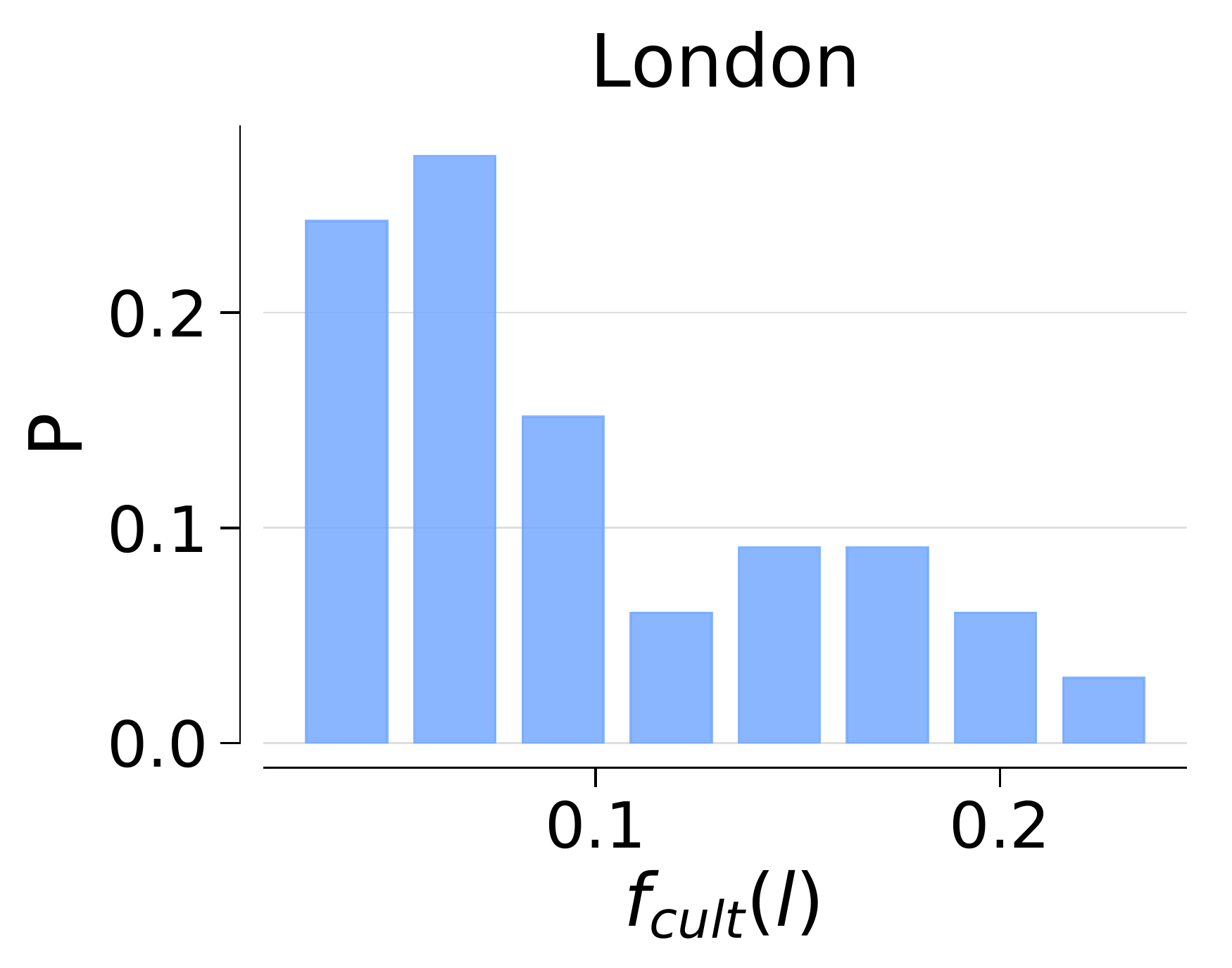} 
	 \includegraphics[width=0.23\textwidth]{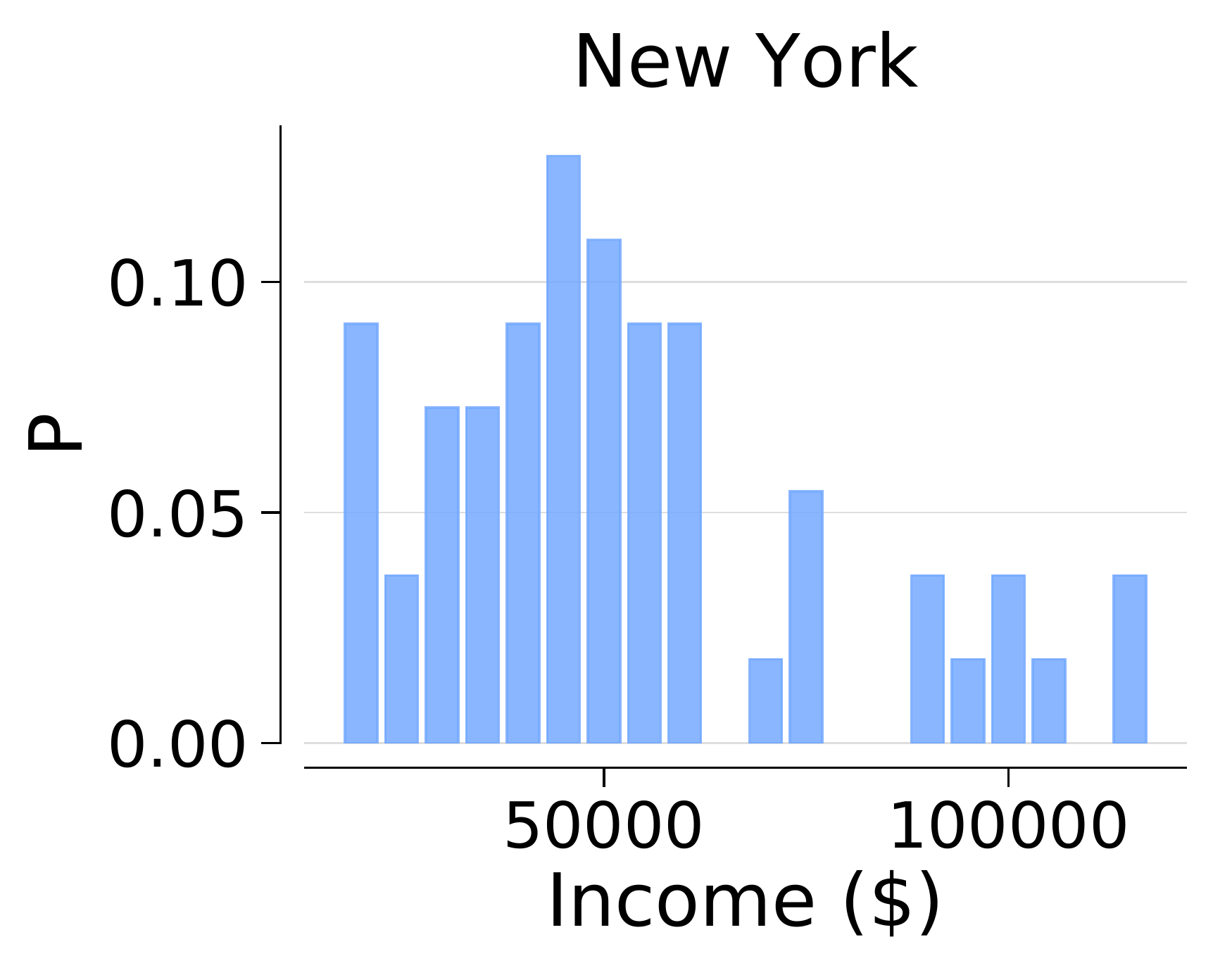}
	 \includegraphics[width=0.23\textwidth]{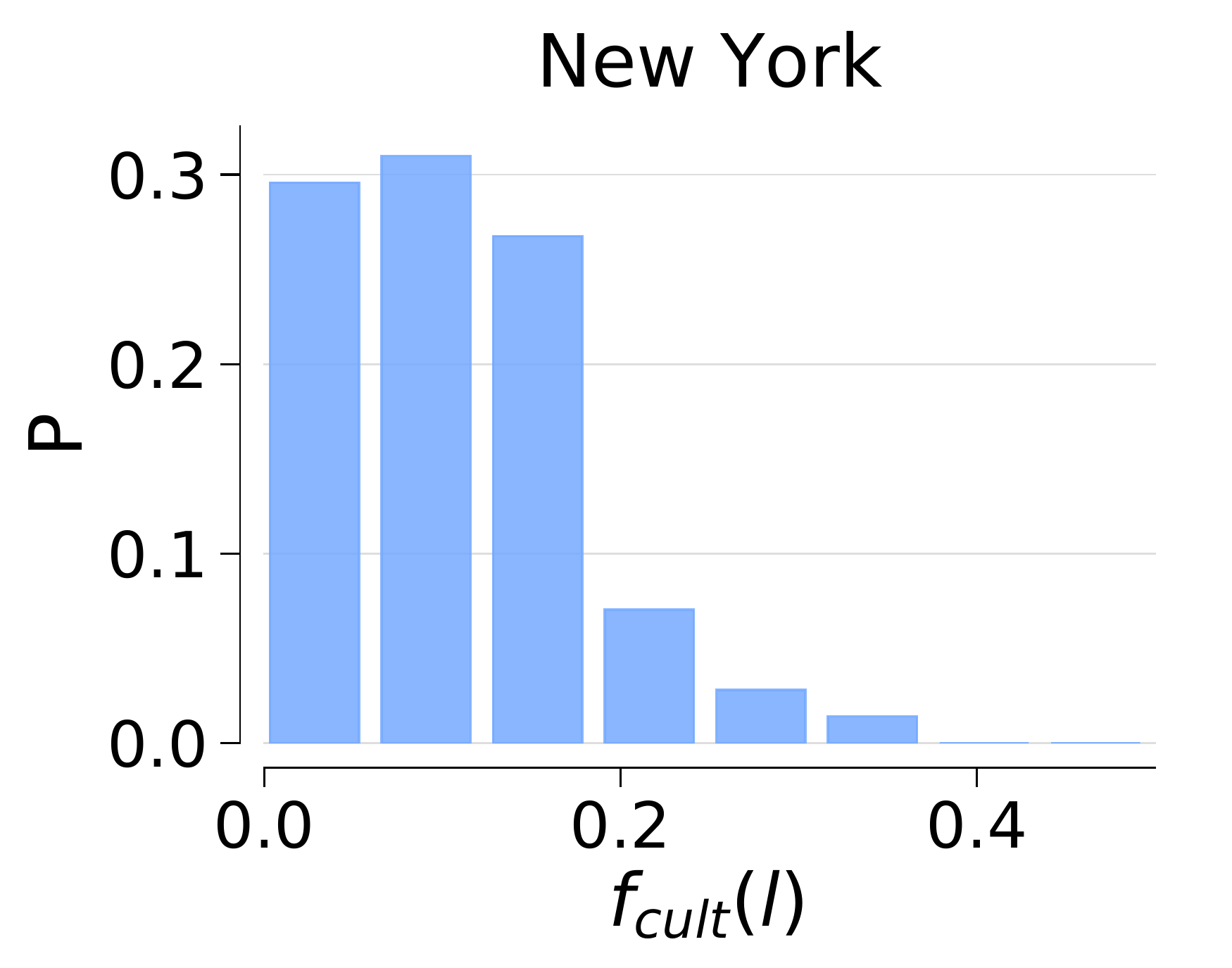}
   \caption{Distributions of variables encoding cultural and economic capital in London and New York.}
   \label{fig:variable_distributions}
\end{figure*}

The goal of this study is to explore how the two forms of capital (cultural and economic) are linked to urban development. To meet that goal, we needed to collect metrics that capture urban development. The Index of Multiple Deprivation (IMD) for London boroughs\footnote{\url{https://data.london.gov.uk/dataset/indices-deprivation-2010}} and the Social Vulnerability Index (SVI) for New York census tracts\footnote{SVI New York:~\url{http://data.beta.nyc/dataset/social-vulnerability-index}} (denoted in the following as dev) had been used as proxies for urban development before. Both are composite measures of deprivation across several domains such as education, barriers to housing, crime, employment, and access to resources.We collected the IMD and SVI values for both years 2010 and 2014. The goal of this study is to explore how the two forms of capital (cultural and economic) are linked to urban development. To capture cultural capital, we processed the pictures continuously from 2007 to 2015 as we have previously specified. Figure~\ref{fig:variable_distributions} shows the frequency distributions of cultural capital in both cities. To estimate economic capital, we gather data about London boroughs’ median income and median house prices released every year from 2007 and 2015, and New York's tracts' average income and median house prices released in 2014 (as only that year was publicly available for New York). Figure~\ref{fig:variable_distributions} shows the frequency distributions of income in London (in terms of British pounds) and in New York (in terms of dollars). To capture urban development, we gather census data reporting the London boroughs' Index of Multiple Deprivation (IMD) released in both 2010 and 2014, and New York tracts' Social Vulnerability Index (SVI) released in both 2010 and 2014. Both reflect urban development as they are composite measures of deprivation across several domains such as education, barriers to housing, crime, employment, and access to resources. Both our taxonomy of urban culture and presence of its terms in each London borough and New York tract are made publicly available under \url{http://goodcitylife.org/cultural-analytics}.

\section{Results}

\subsection{Urban development and the forms of capital}

\begin{figure}[t!]
    \centering
    \includegraphics[width=0.47\textwidth]{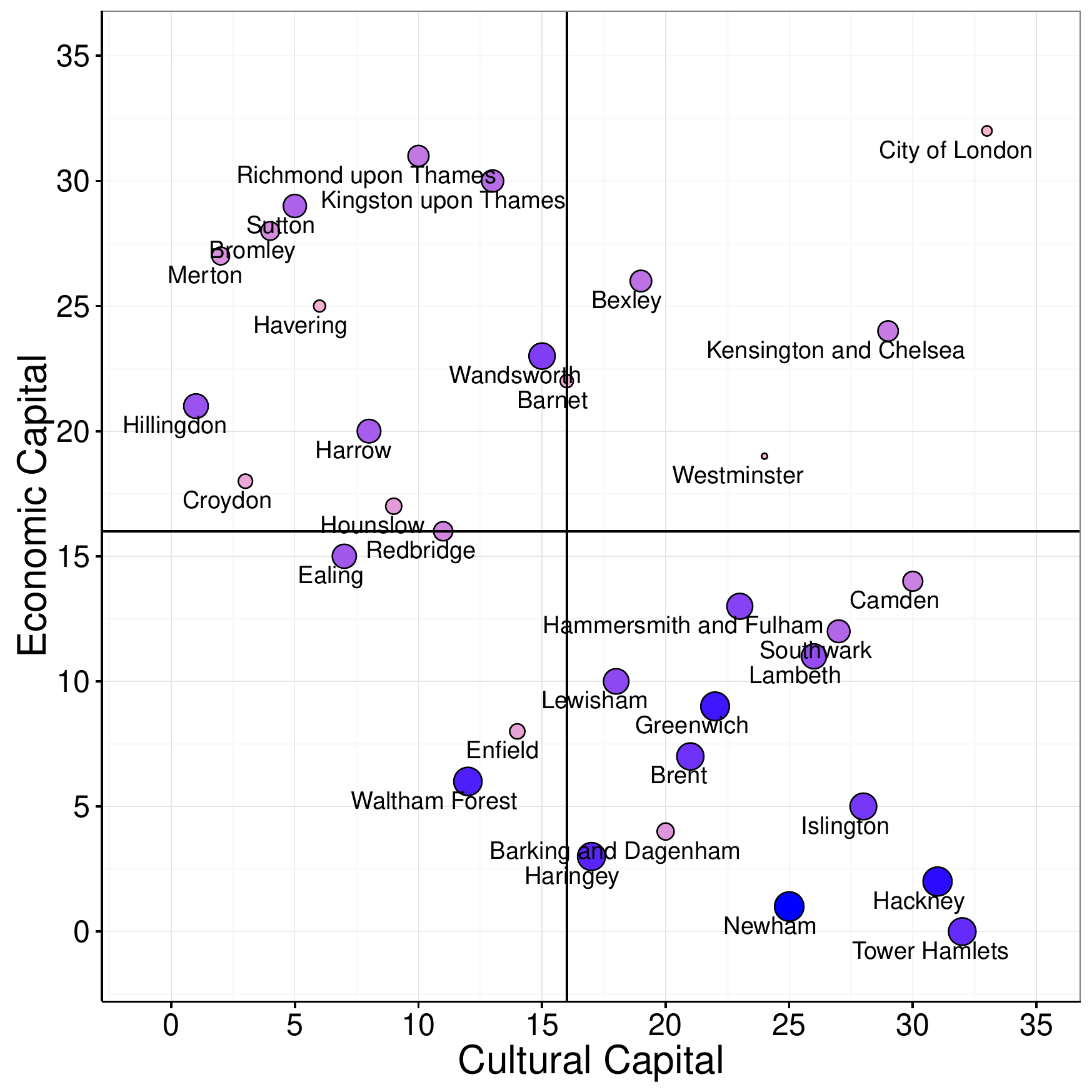}
		\includegraphics[width=0.47\textwidth]{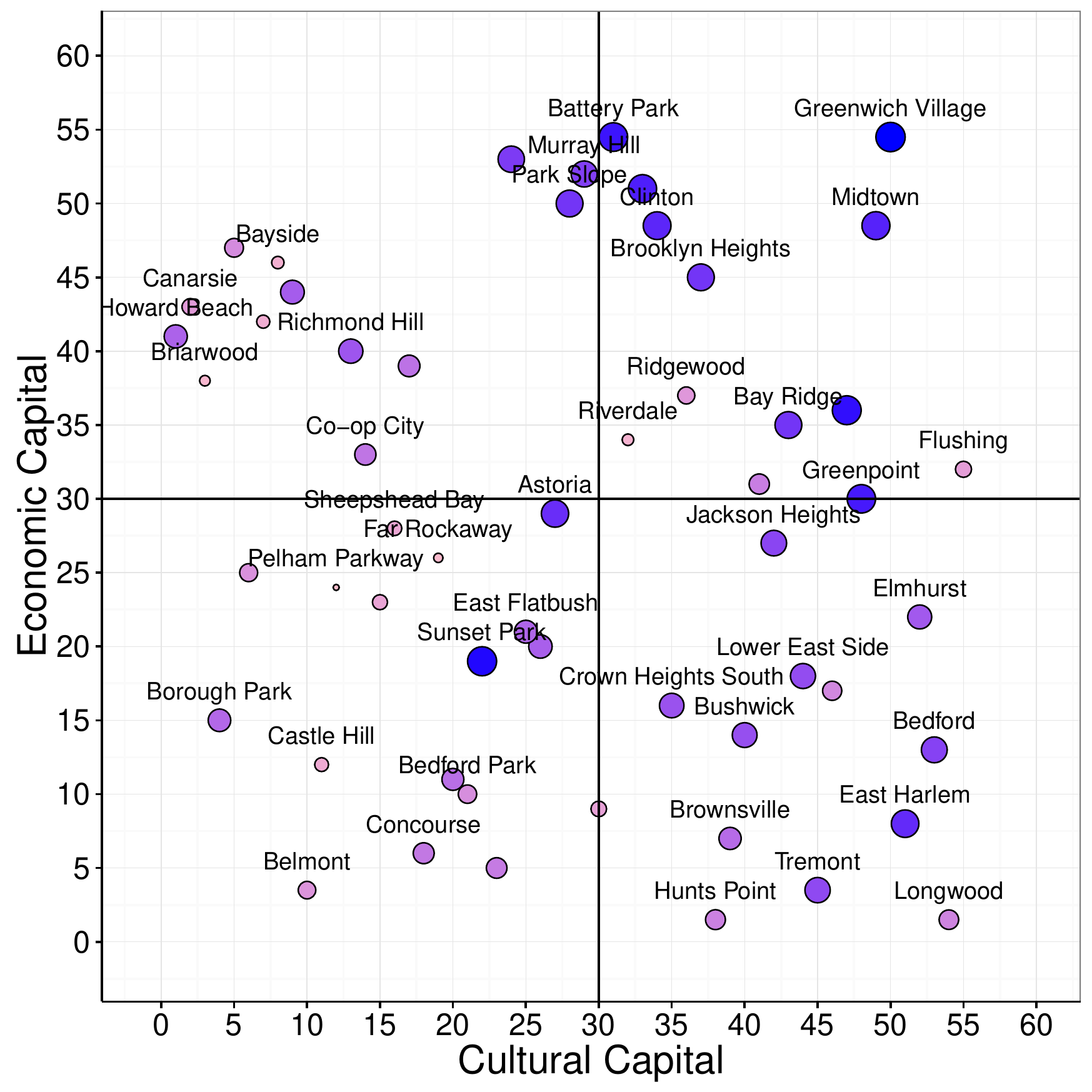}
    \caption{Ranked cultural and economic capital in 2010. The size of each node represents the change in development score (IMD for London and SVI in New York) between 2010-2015. The darker and larger a node is, the more improvement it has had over the five year period.}
    \label{fig:cap}
\end{figure}

Following the framework defined by Bourdieu in the context of social class and drawing an analogy between social class and urban development, we ask if for neighborhoods, much like for people, cultural capital leads to positive development. As Bourdieu himself argued, prosperity cannot be fully explained by economic capital alone. We therefore considered both cultural capital and economic capital of neighborhoods in 2010 and checked to what extent they predicted urban development (IMD in London and SVI in New York) five years later---at the beginning of 2015---through the following linear regression:
\begin{equation}
 dev = \alpha + \beta_1 \cdot capital_{cult}  + \beta_2 \cdot capital_{econ}  
 \label{eq:model}
\end{equation}
For London, cultural and economic capital in 2010 are used to predict urban development in 2015. For New York, where less granular data is available, the average income in the period 2010-2014 is used instead. We used ordinary least squares regression as a method to fit the data. Cultural and economic capital measured in 2010 are strong predictors of the development of an area in 2015 in both cities (Table~\ref{tab1}). In New York, the development score is relatively well explained by economic capital alone ($R^2=0.75$), whereas in London cultural capital also plays an important role in the prediction (both regression coefficients are significant). Nevertheless, when it comes to changes in development scores from 2010 to 2015 (referred to as $\Delta dev$), we find that both types of capital have comparable roles in both cities. This suggests that improvement in neighborhoods is a function of both economic capital and cultural capital.  This is visually confirmed in Figure~\ref{fig:cap} where each dot corresponds to a neighborhood, its position depends on the two values of capital for that neighborhood, and its size reflects a positive change in development  over the five years under study. As opposed to what happens in London, in New York,  improvements are more significant for already economically prosperous neighborhoods. Despite this difference, cultural capital still remains a powerful currency in both cities: positive changes in development scores are higher along the cultural capital axis in both cities. When controlling for Flickr penetration (z-score of number of tags in the neighborhood) the results change only slightly, with a relative increase in $R^2$ between $-1\%$ and $+12\%$, meaning that the actual cultural content---and not the volume of all types of photos---is predictive of development.

\begin{table}[b]
\small
\begin{center}
\begin{tabular}{c|ccc|c}
& \multicolumn{3}{c|}{Regression coefficients} & $R^2$\\
                   & $\alpha$ & $capital_{cult}$ & $capital_{econ}$ & \\
\hline
$dev_{lon}$ & -0.51 & 3.4$^{***}$  & -4.5$^{***}$  & 0.65 \\
$dev_{ny}$  &   0.0 & -0.02   & -0.87$^{***}$ & 0.75 \\
\hline
$\Delta dev_{lon}$ & -0.18 &  2.26$^{**}$ & -3.91$^{***}$ & 0.38 \\
$\Delta dev_{ny}$  &  0.02 & 0.381$^{**}$ & 0.46$^{***}$  & 0.33
\end{tabular}
\end{center}
\caption{Linear regression to model urban development $dev$ (at the beginning of 2015) in terms of economic capital and social capital. Regression coefficients, goodness of fit ($^{**}$p\textless 0.001; $^{***}$p\textless0), and the coefficient of determination $R^2$ are shown. $\Delta dev$ is the difference between the development measured in 2015 and that measured in 2010.}
\label{tab1}
\end{table}

\subsection{Digital cultural profiles and development}

\begin{figure}[t!]
    \centering
    \includegraphics[width=0.49\textwidth]{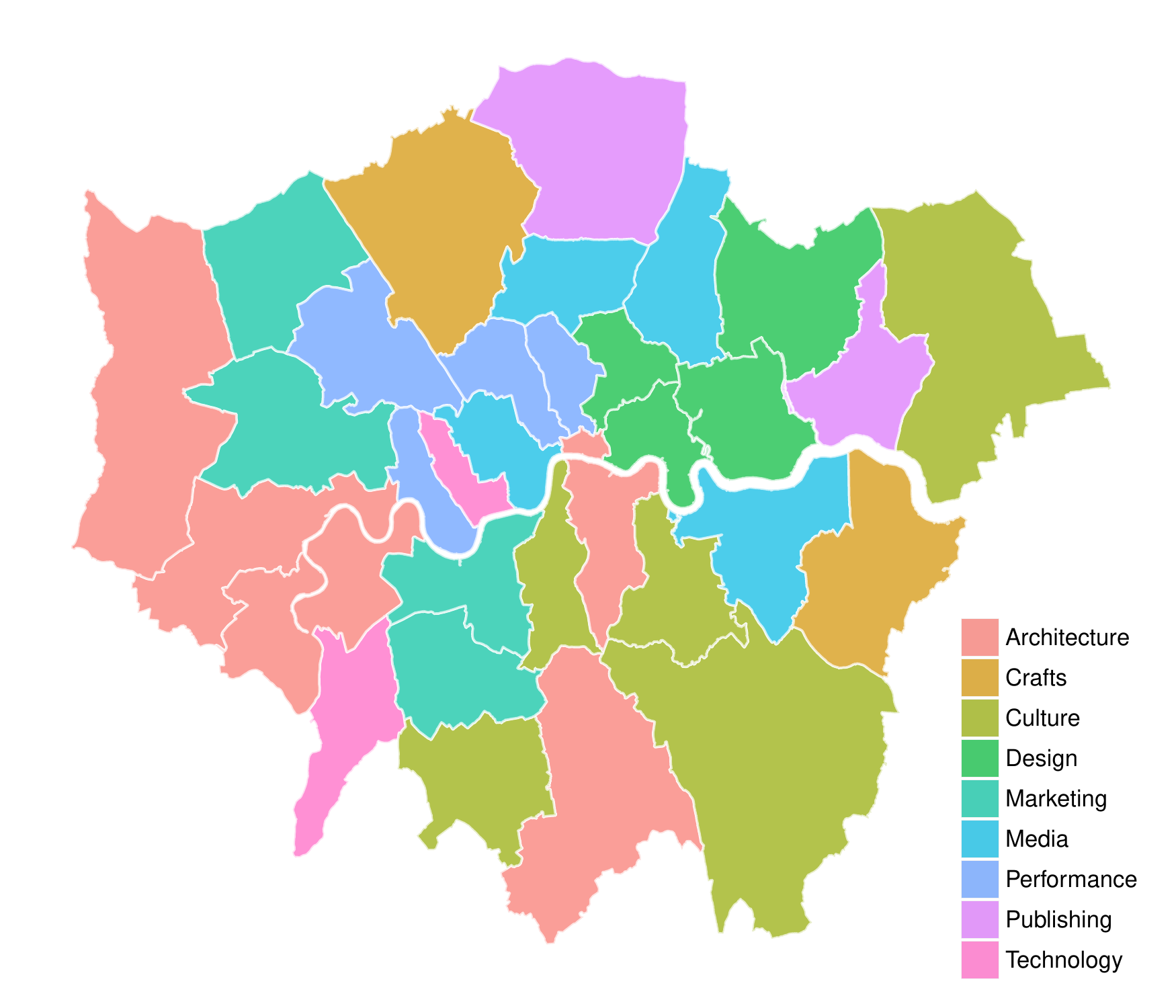}
		\includegraphics[width=0.49\textwidth]{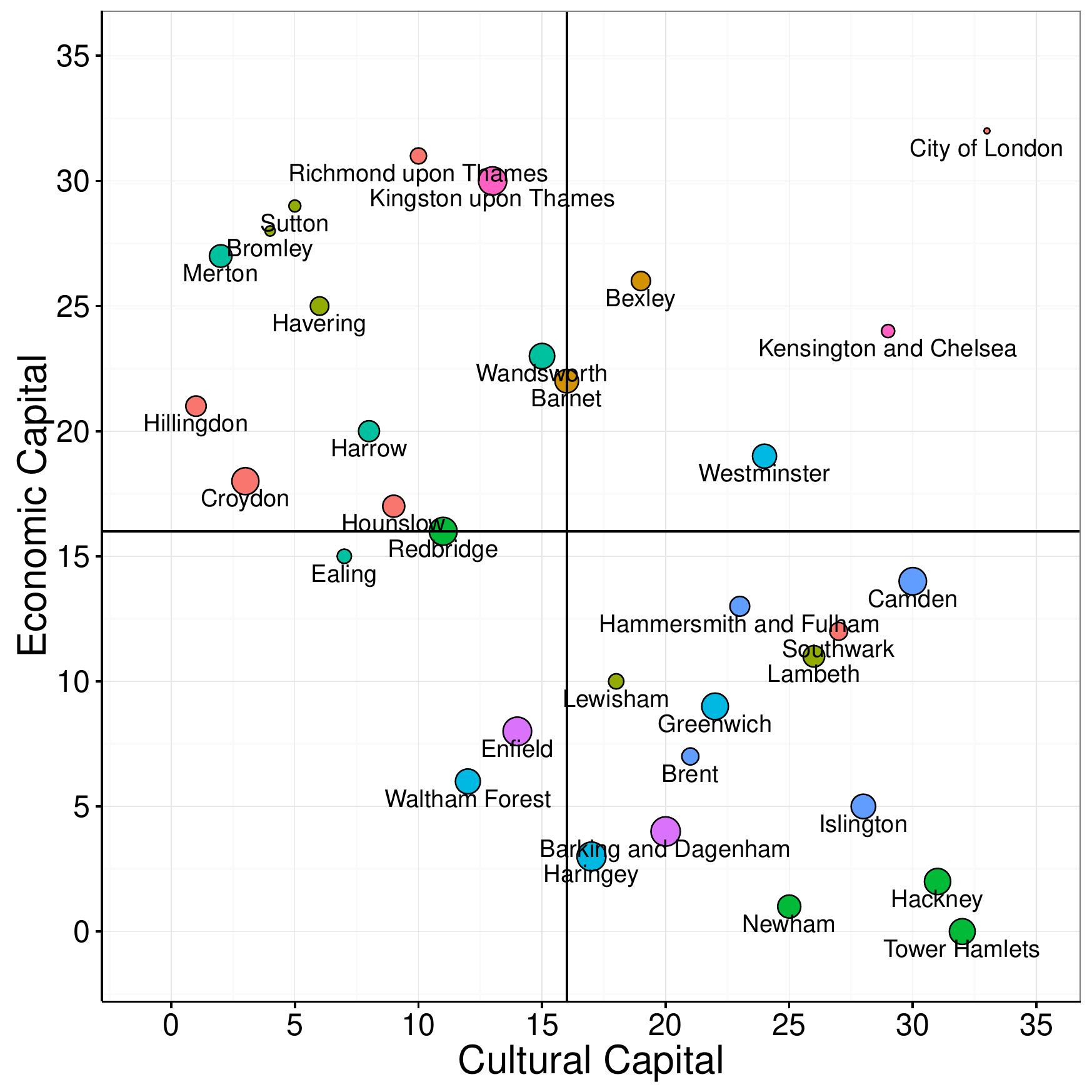}
		\\
		\includegraphics[width=0.49\textwidth]{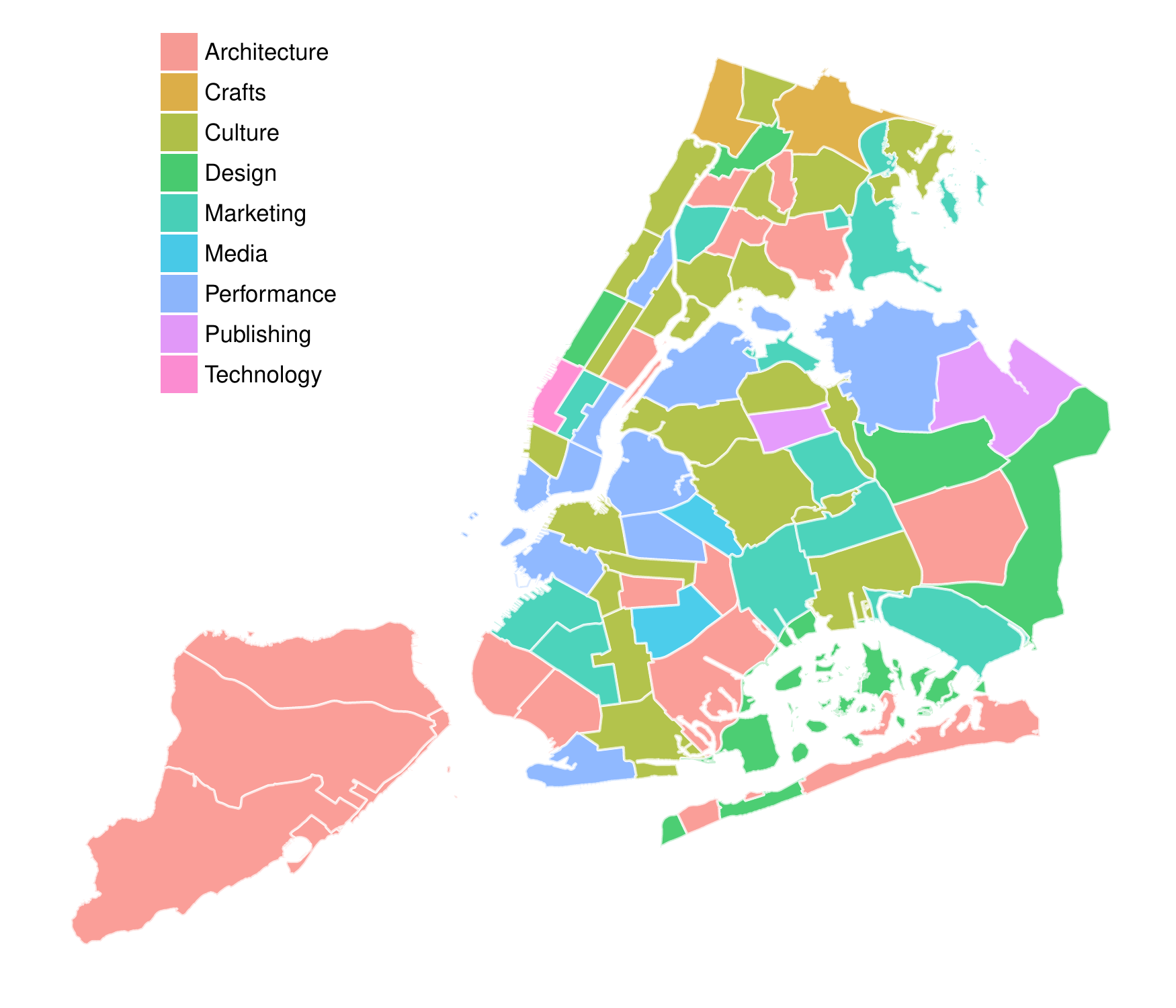}
		\includegraphics[width=0.49\textwidth]{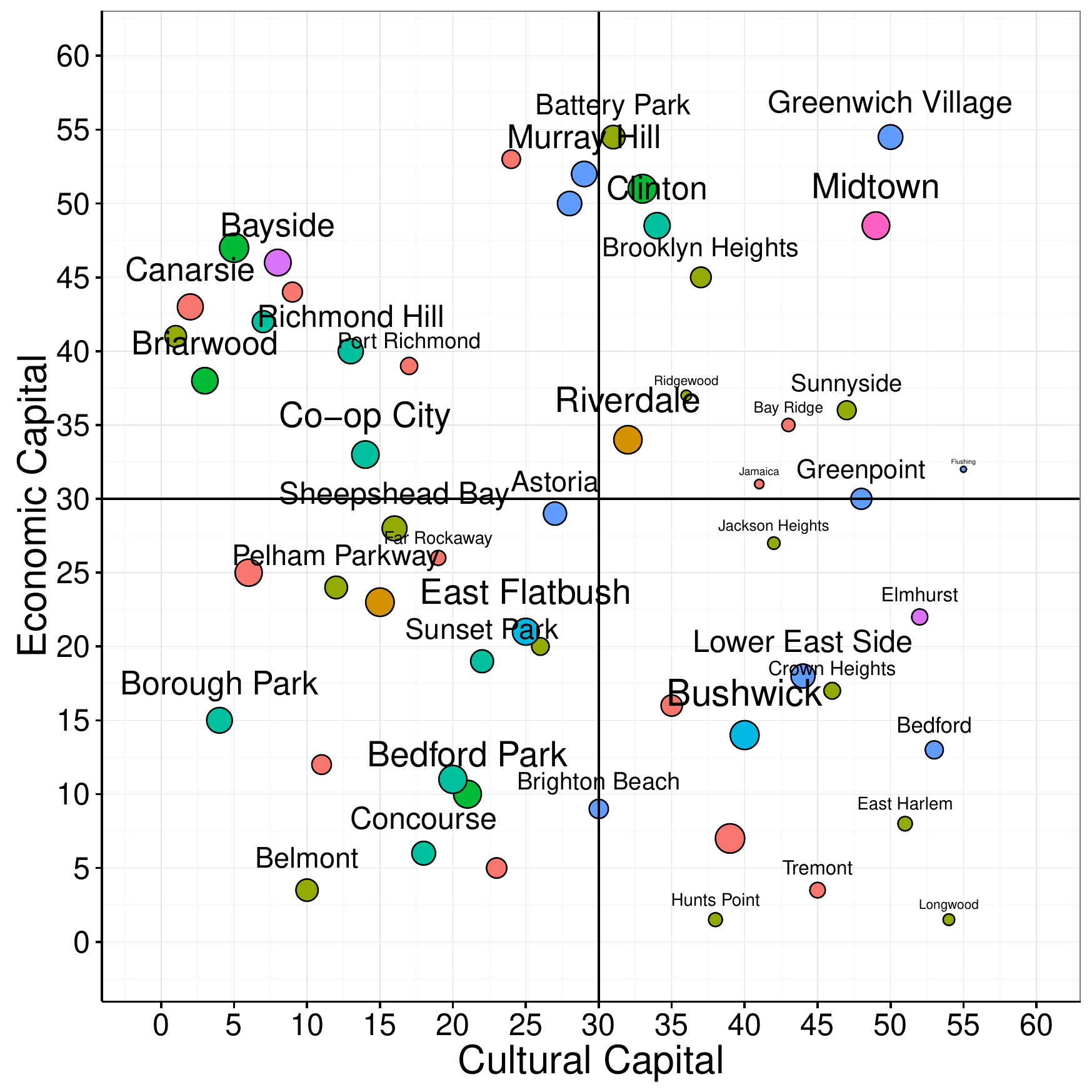}
    \caption{The cultural specialisation of neighbourhoods in London (top) and New York (bottom). On the left, the maps of the most representative cultural assets in different neighbourhoods; on the right, quadrants that relate cultural capital, economic capital, cultural specialisation (color of dots), and cultural diversity (size of dots).}
    \label{fig:spec}
\end{figure}

\begin{figure}[t!]
    \centering
    \includegraphics[width=0.49\textwidth]{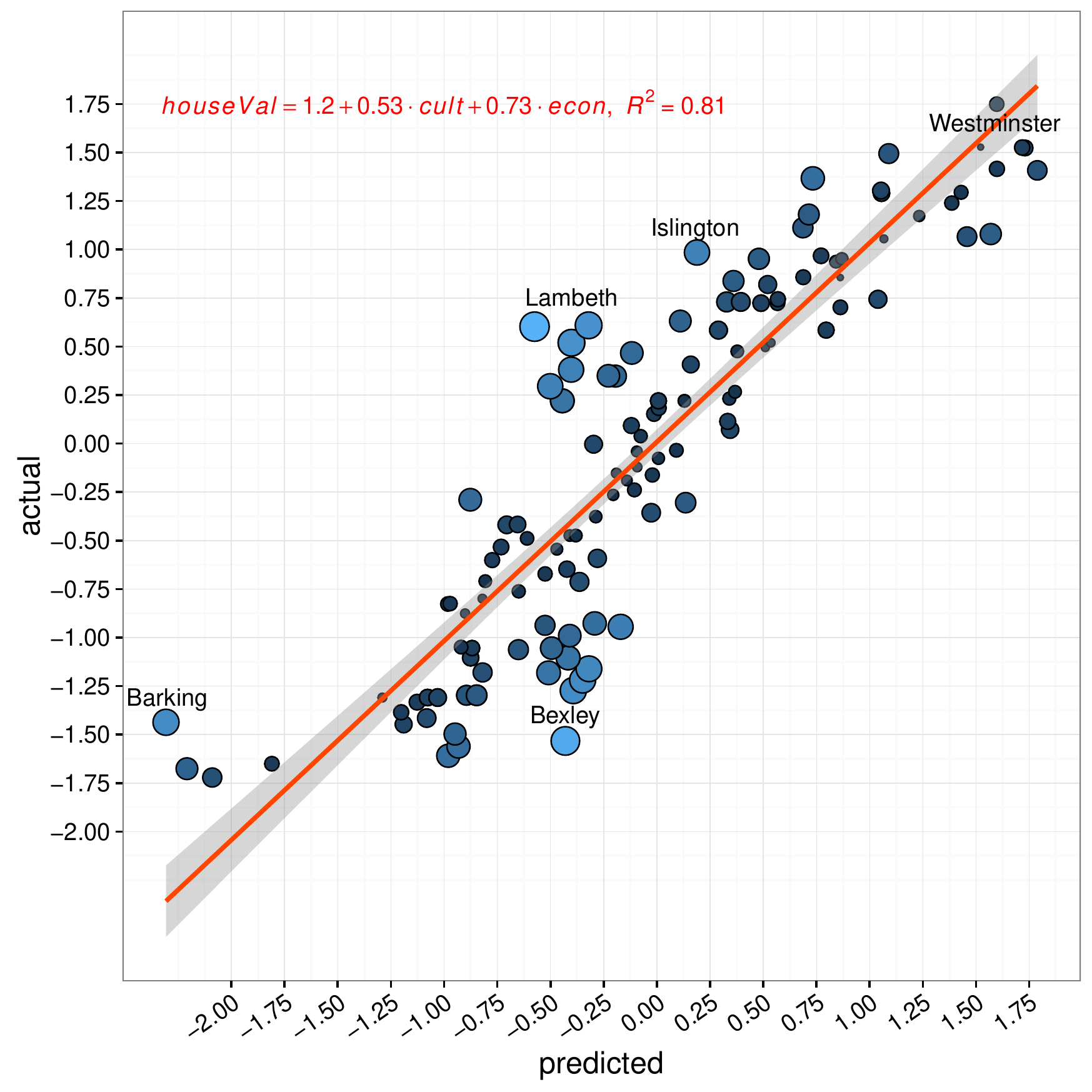}
		\includegraphics[width=0.49\textwidth]{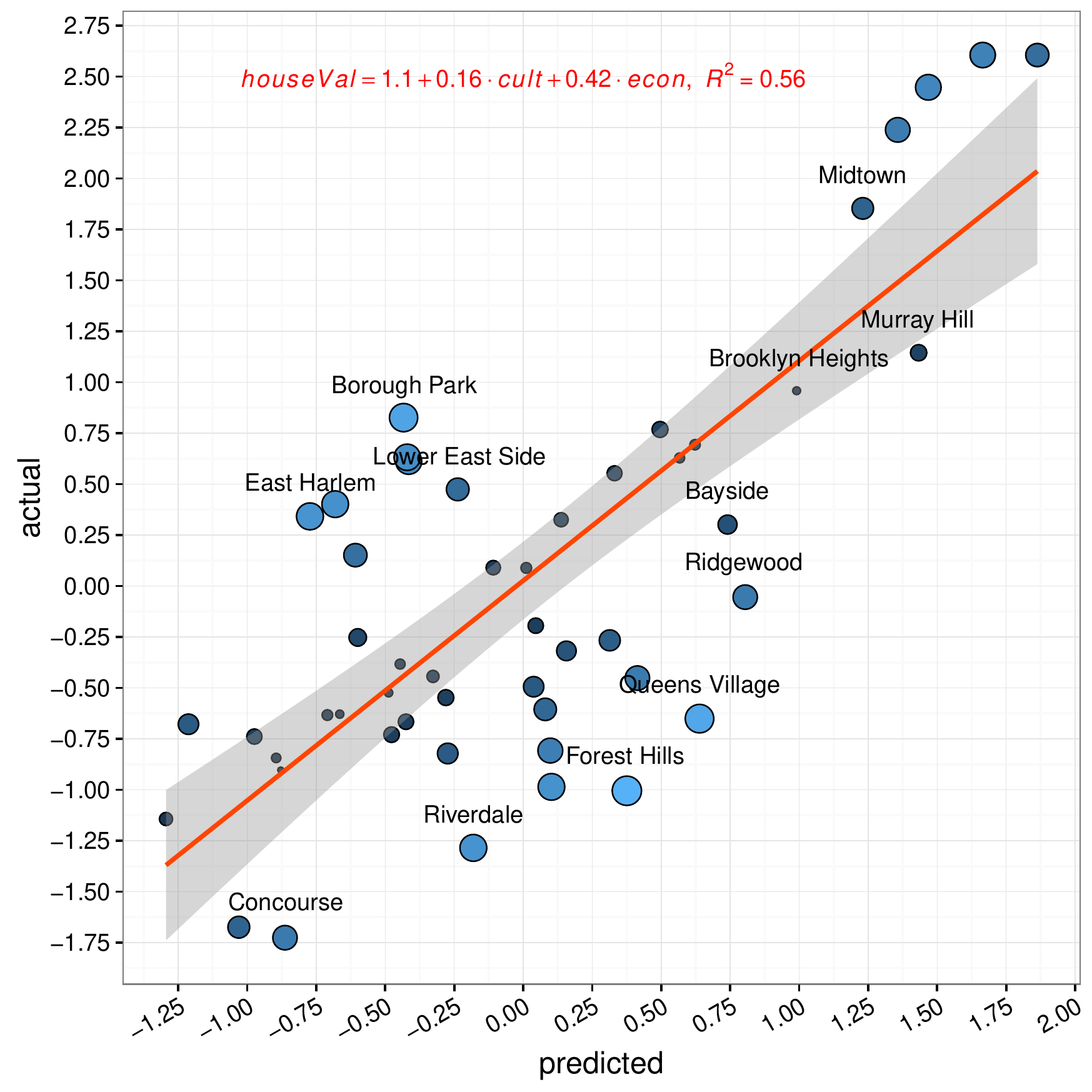}
    \caption{Linear regression results for housing price $z$-scores across neighbourhoods over the period 2010-2015. The regression line is shown in red and the shaded area around it represents the limits of the 95\% confidence interval.}
    \label{fig:house}
\end{figure}

To see in what kind of activities cultural capital translates into, we explored which  \textit{type} of culture is consumed in different neighborhoods. We measured the cultural characteristics of a location in terms of the nine top-level dimensions of our taxonomy.  We mapped the cultural specialization of neighborhoods ($special_{cult}$, defined in Equation~\ref{eq:specialisation}) in Figure~\ref{fig:spec} (left panels). In both cities, `Performance Arts' appears in central areas, and `Architecture' is predominant in either central areas and peripheral ones.  As opposed to New York, London deviates from this typical pattern at times: East London specializes in `Design', and West London specializes in `Performance Arts'  and `Marketing'.

To place these observations in context, we drew again the quadrant of economic capital \emph{vs.} cultural capital (Figure~\ref{fig:spec}, right panels), but, this time, the color of a node reflects the corresponding location's specialization, and its size reflects the location's cultural diversity ($diversity_{cult}$ in Equation~\ref{eq:diversity}). In both cities, neighborhoods with high cultural capital do specialize in `Performance Arts' (in London, they do specialize  in `Design' and, to a lesser extent, in `Publishing and Media' too). Also, neighborhoods with increasing urban development tend to be high not only in cultural capital but also in cultural diversity. The observation that higher urban development is associated with cultural diversity is in line with what urbanists have claimed to be the main driver of neighborhood prosperity: having diverse industries in geographical proximity~\cite{glaeser2011triumph,jacobs1969economy}. Indeed, by adding entropy as a regressor in our model in Equation~\ref{eq:model}, we improved our goodness of fit to $R^2=0.41$ for $\Delta dev_{lon}$ and $R^2=0.71$ for $dev_{lon}$ for London ($8\%$ and $9\%$ improvement, respectively). 

\subsection{Housing and cultural capital}
One of the main concerns that developing neighborhoods face is  increasing house prices.  If urban development is analogous to social mobility, then the house value of a neighbourhood can be compared to social class in Bourdieu's terms. Therefore, similar to what we did for urban development, we used a linear regression to predict house prices from the cultural and economic capital with the following regression:
\begin{equation}
 house\_price = \alpha + \beta_1 \cdot capital_{cult} + \beta_2 \cdot capital_{econ} 
\label{eq:model2}
\end{equation}
For London, cultural and economic capital in 2010 are used to predict the housing prince in 2015. For New York, where less granular data is available, the cultural capital in 2010 and the economic capital in the period 2010-2014 is used to predict the average house price in the period 2010-2014. The results suggest that, again, the ability to predict and explain house prices is not a purely economic matter (Figure~\ref{fig:house}). Both forms of capital play a significant role in the model ($\beta_1 = 0.53, \beta_2=0.73$) with an $R^2$ of $0.81$ in London, and $0.56$ in New York.
Naturally, people choose to live in areas they can afford and therefore  economic capital still plays a fundamental role in explaining housing prices, however, an economic regressor alone achieves an $R^2 = 0.53$ in London, and $R^2=0.48$ in New York, showing the importance of cultural capital. By then re-computing a regression for each of the 9 top-level categories of cultural capital (Table~\ref{tabcc}), we explored whether a specific  \emph{type} of cultural capital is associated with increasing house prices.  In New York, `\emph{Publishing}'  (e.g., content labeled as \emph{newspaper} or \emph{books})  is most indicative of increasing housing prices, and a linear model for predicting house prices with it alone outperforms the composite cultural capital model ($R^2$=0.59). In London, `\emph{Technology}' is associated with increasing house prices ($R^2 = 0.79$), but a model with it alone does not outperform the composite cultural capital model. 

Overall, taken together, the previous results suggest that even though several economic and geographical factors impact house prices---such as property type or size, which we do not consider here---cultural capital alone holds a considerable explanatory power. 

\begin{table}
\begin{center}
\begin{tabular}{ l l l}
      & London & NY\\ \hline
Architecture & 0.71 & 0.52\\
Crafts & 0.63 & 0.55\\ 
Culture & 0.64 & 0.58\\ 
Design & 0.62 & 0.57\\ 
Marketing & 0.62 & 0.5\\ 
Media & 0.63 & 0.51\\ 
Performance & 0.64 &0.52 \\ 
Publishing & 0.64 & \textbf{0.59}\\ 
Technology & \textbf{0.79} & 0.51\\ 
\end{tabular}
\end{center}
\caption{$R^2$ coefficients for different types of cultural activities in predicting housing prices.}
\label{tabcc}
\end{table}

\subsection{Generating and spending cultural capital}

\begin{figure}[t!]
    \centering
		\includegraphics[width=0.65\textwidth]{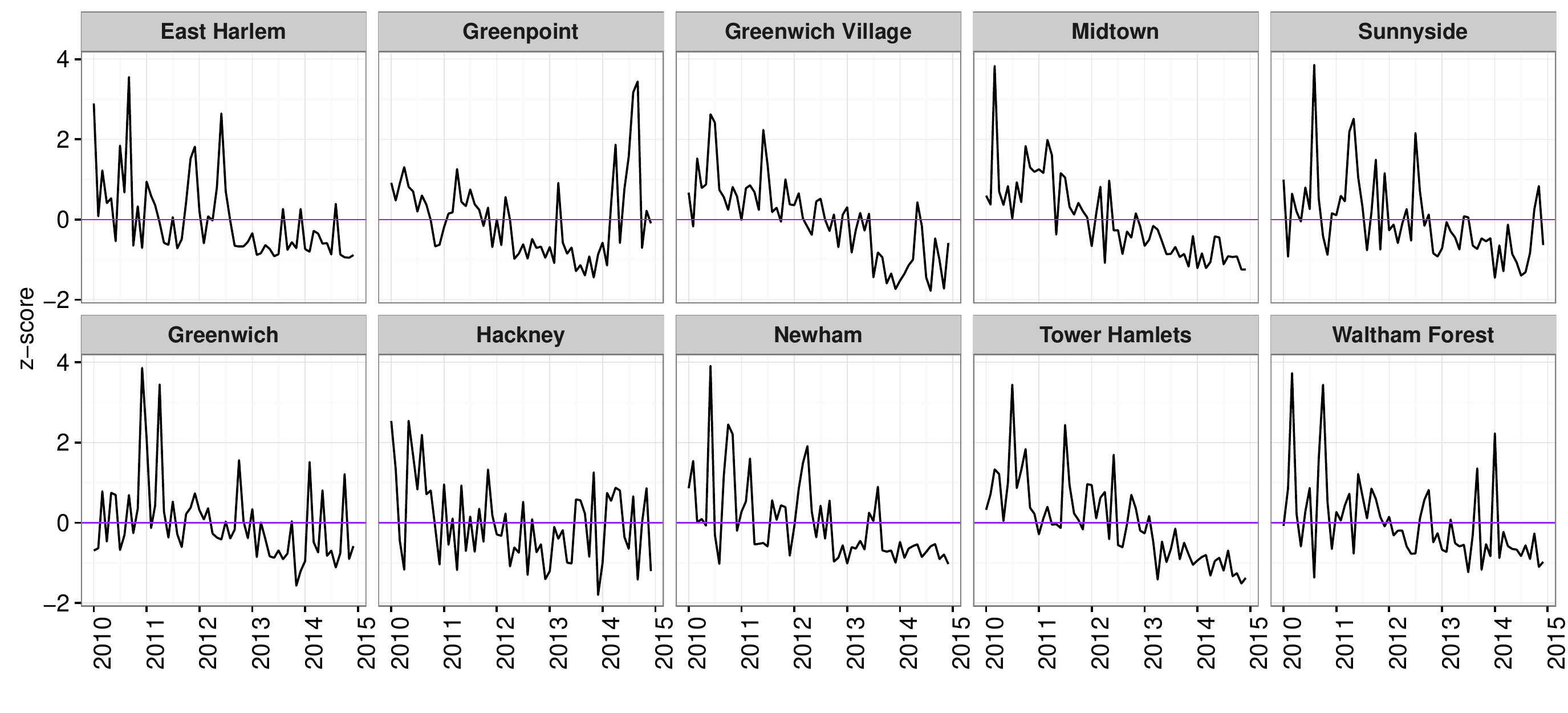}
    \includegraphics[width=0.30\textwidth]{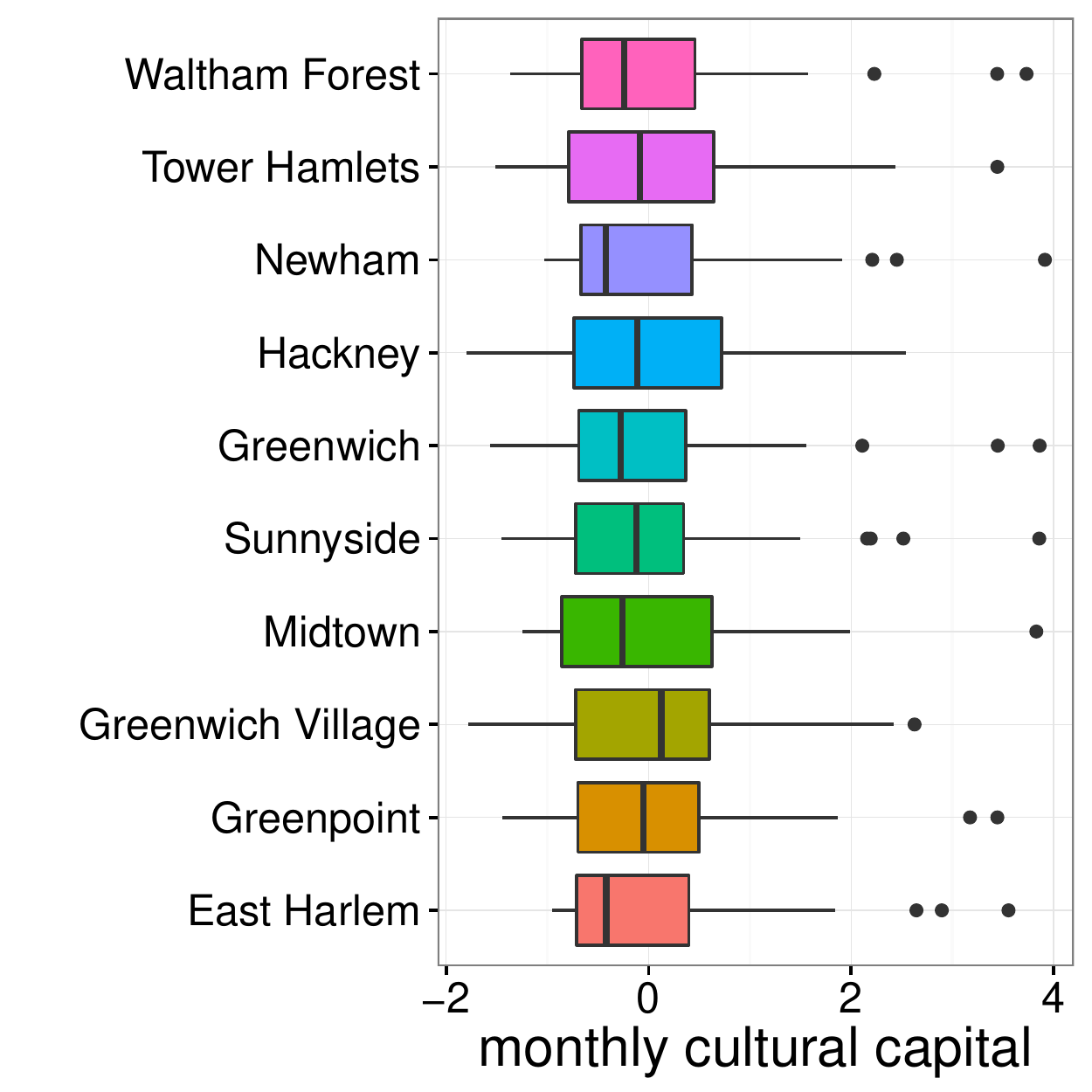}
    \caption{Cultural capital for the five neighborhoods in London and New York that most improved over the five year period (2010-2015). Left: inter-borough $z$-scores of cultural capital on a monthly basis shows divergences from the mean (purple line). Right: Distribution of cultural capital monthly values across neighborhoods (outliers are events with considerably higher cultural capital).}
\label{fig:events}
\end{figure}

We have seen that cultural capital is associated with socio-economic development and increasing house prices.  One might now wonder how cultural capital is generated. Since one  simple way is through cultural events, we set out to detect such events in our  our digital data. To detect peaks in the fluctuation of the cultural capital that might correspond to key cultural events, we measured the cultural capital of a neighborhood on a running monthly basis and compared it to the expected value in that neighborhood. More specifically, we computed the $z$-score of the fraction of cultural content at month $t \in [0,T]$ using the average and standard deviation of the fraction measured in all months ($0$ to $T$) at location $l$:
\begin{equation}
capital^t_{cult}(l) =  \frac{f^t_{cult}(l) - \mu(f^{[0-T]}_{cult}(l))}{\sigma(f^{[0-T]}_{cult}(l))}.
\end{equation}
Figure~\ref{fig:events} (left) shows the variation of the cultural capital over time for the five neighborhoods in London and New York that had the highest variation of urban development ($\Delta dev$) between 2010 and 2015. Peaks and falls are easy to see, if contrasted to the horizontal line (which is the neighborhood's (typical) mean cultural capital). For each of the top outliers in the neighborhoods in Figure~\ref{fig:events}, we identified the exact event that took place. In Table~\ref{tab:events}, we see that a variety of cultural events were indeed at the heart of changing and enhancing the reputation of specific places in both cities. 

\begin{figure}[t!]
    \centering
		\includegraphics[width=0.99\textwidth]{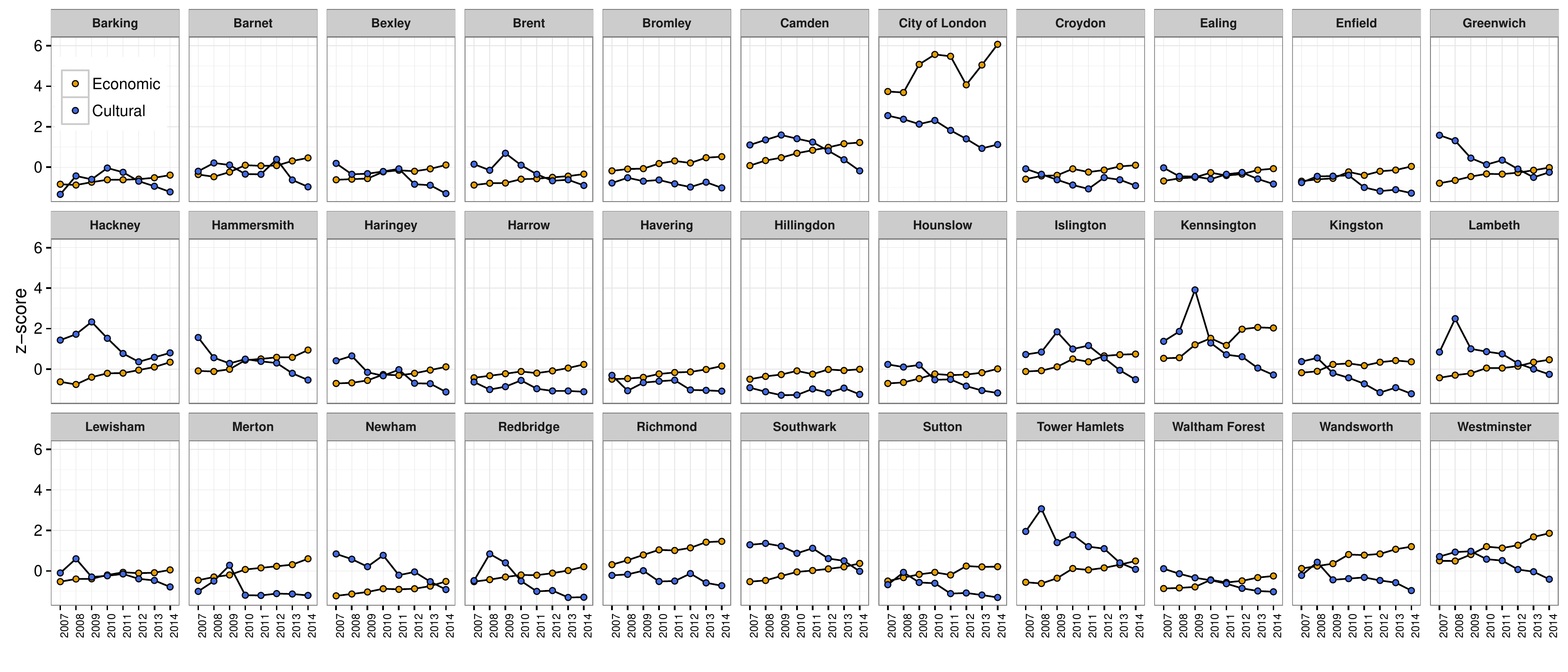}
    \caption{Evolution of cultural and economic capital for different London boroughs in the period 2007-2014. Because the values are z-score normalized they are comparable: a borough with an economic capital higher than the social capital (or viceversa) indicates means that, relative to all other neighborhoods, that borough is better in terms of its economic rather than cultural status. Curves crossing indicate that one type of capital becomes more prominent than the other, relatively to all other neighborhoods.}
\label{fig:gentrification}
\end{figure}

Based on a wider temporal analysis in London (Figure~\ref{fig:gentrification}), one can see that cultural capital translates into economic capital in a few years. Areas subject to cultural revitalization eventually gentrify~\cite{zukin1987gentrification,florida2003cities}.

\begin{table*}[h!]
\centering
\tiny
  \begin{tabular}{ | l | L{2cm} | L{2cm} | l | l | l | l | C{2cm}  }
  \hline
  $date$ & $event$ & $borough$ & $photos/users$ & $cult$ & $category$ & $image$\\ \hline
    \hline
  6 Mar 2010 & East London murals walk & Hackney & 4.6 & 2.1 &  Art & \includegraphics[scale=0.1]{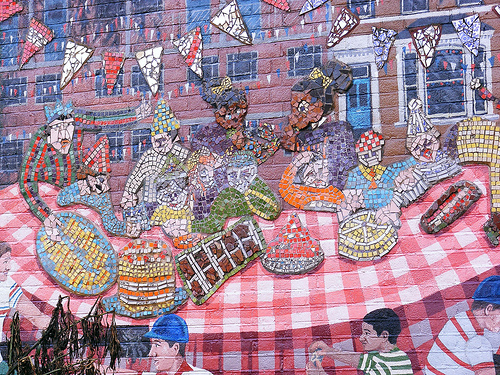}\\ 
      \hline
  7 Dec 2010 & Suede music concert, O2 Arena & Greenwich & 20.4 & 3.45 & Performance & \includegraphics[scale=0.1]{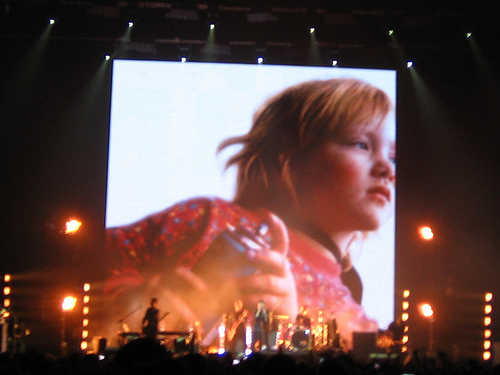}\\ 
        \hline
  5 Jun 2010 & Miss Southern Africa UK & Newham & 159.6 & 3.9 & Culture & \includegraphics[scale=0.1]{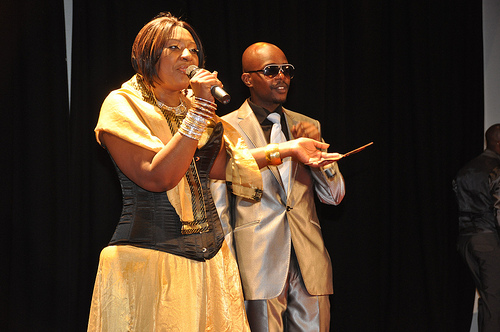}\\ 
          \hline
  24 Jul 2010 & High Voltage Festival & Tower Hamlets & 27.8 &  3.44 & Performance & \includegraphics[scale=0.1]{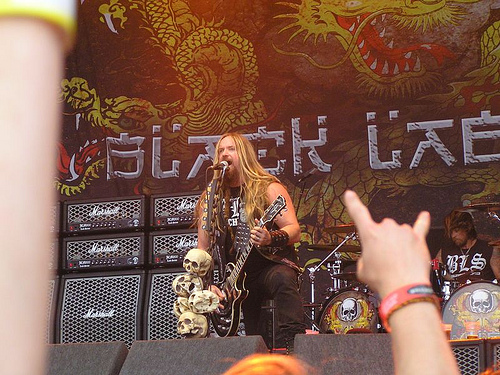}\\ 
            \hline
 17 Mar 2010 & Boyz Magazine Cabaret & Waltham Forest & 27 & 3.7 & Culture & \includegraphics[scale=0.45]{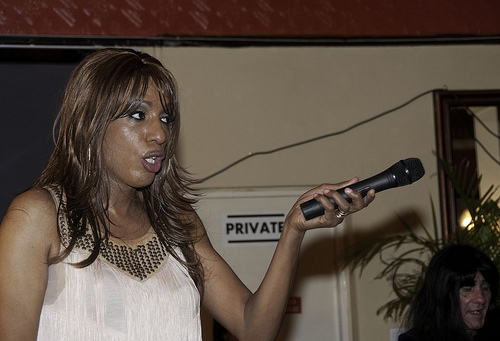}\\ 
    \hline \hline
 21 Aug 2010 & PS1 MoMa Young Architect's Program & Sunnyside & 40.6 & 3.85 & Culture &\includegraphics[scale=0.15]{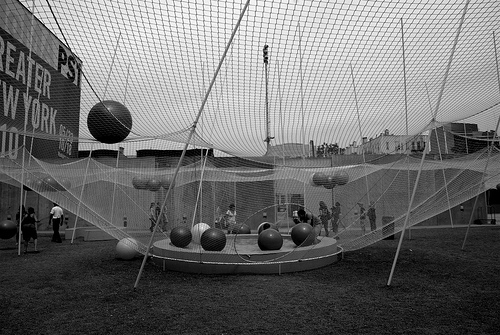}\\ 
    \hline 
     5 Jun 2010 & LGBT Community Centre figure drawing & Greenwich Village & 21.16 & 2.62 & Culture &\includegraphics[scale=0.25]{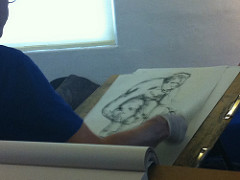}\\ 
    \hline 
        13 Sep 2014 & 13th Annual Johnny Heff Tribute & Greenpoint & 450 & 3.44 & Performance & \includegraphics[scale=0.25]{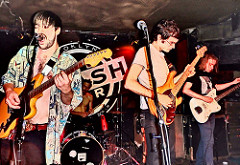}\\ 
    \hline 
      20 Mar 2010 & 17th Original GLBT Expo & Midtown & 52.6 & 3.82 & Culture & \includegraphics[scale=0.25]{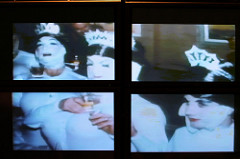}\\ 
    \hline 
       5 Sep 2010 & Electric Zoo Festival & East Harlem & 46.4 & 3.55 & Culture & \includegraphics[scale=0.25]{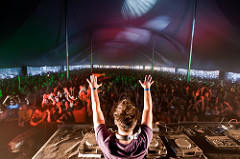}\\ 
    \hline 
  \end{tabular}
  \caption{Summary of top events in the five neighborhoods of London and New York that  most improved from 2010 to 2015. For each event, the table reports its date, name, location, number of users involved in it, the change it caused  in cultural capital in terms of deviation from the mean ($cult$), the most frequently occurring cultural category, and a representative image.}
 \label{tab:events}
\end{table*}

\section{Conclusion}

Culture pays. That is not always obvious for policy makers. ``When budgets have come under pressure, there has been a tendency for arts and culture to be viewed as `nice to have', rather than a necessity''~\cite{henley2016arts}. Culture surely  comes with \emph{intrinsic} benefits: it  opens our minds to new emotional  experiences, and enriches our lives. But, as we have shown, it also comes with \emph{extrinsic} benefits: it is a catalyst for positive change and growth in neighborhoods. We have found that the neighborhoods in London and New York that experience the greatest growth are those with high cultural capital. The production of these findings relies on a new way of quantifying cultural capital that is based on the definition of the first taxonomy of  culture (which is far more comprehensive than official classifications of cultural activities) and on the mining of digital data such as picture tags  (which has made it possible to perform cultural studies at an unprecedented scale, contributing to the emergence of a new research field called `\emph{cultural analytics}'~\cite{manovich2009cultural}). Despite being geographically biased, picture data has been valuable not only for observing neighborhood growth and identifying ``up-and-coming'' areas but also for predicting house prices.

Culture pays, but only up to a point. As Bourdieu argued, cultural inequality widens and legitimizes economic inequality~\cite{grenfell2014pierre}. As such, culture---which  powers the growth of cities---also causes their distressing challenges: gentrification, unaffordability, and inequality~\cite{florida2017new}. A sustainable approach to cultural investments might pay dividends but requires sensitivity to the needs of local communities. 

\bibliographystyle{ACM-Reference-Format}

\end{document}